%% file: HoloGravAnomal.tex
\documentclass{JHEP3}
\usepackage[english]{babel}
\usepackage{amssymb}
\usepackage{amsfonts}
\usepackage{amsmath}
\usepackage{graphicx}
\usepackage{epsfig}
\usepackage{cite}
\def\erf#1{(\ref{#1})} 
\newcommand{\cA}{{\cal A}}  \newcommand{\cB}{{\cal B}}

\newcommand{\cO}{{\cal O}}  
  
  \newcommand{\cT}{{\cal T}}
  \newcommand{\cV}{{\cal V}}

\newcommand{\be}{\begin{equation}} \newcommand{\ee}{\end{equation}}
\newcommand{\bea}{\begin{eqnarray}} \newcommand{\eea}{\end{eqnarray}}
\newcommand{\beann}{\begin{eqnarray*}}  \newcommand{\eeann}{\end{eqnarray*}}
\newcommand{\bfig}{\begin{figure}} \newcommand{\efig}{\end{figure}}
\newcommand{\ba}{\begin{array}} \newcommand{\ea}{\end{array}}
\newcommand{\bcen}{\begin{center}} \newcommand{\ecen}{\end{center}}
\newcommand{\btab}{\begin{tabular}} \newcommand{\etab}{\end{tabular}}
\newcommand{\nn}{\nonumber}
\newcommand{\dd}{{\rm d}}
\newcommand{\e}{{\rm e}}

\newtheorem{Proposition}{Proposition}[section]

\newtheorem{Theorem}{Theorem}[section]
\newtheorem{Lemma}{Lemma}[section]
\newtheorem{Corrolary}{Corrolary}[section]

\newcommand{\bp}{\begin{Proposition}}	\newcommand{\ep}{\end{Proposition}}
\newcommand{\bt}{\begin{Theorem}}	\newcommand{\et}{\end{Theorem}}
\newcommand{\bl}{\begin{Lemma}}		\newcommand{\el}{\end{Lemma}}
\newcommand{\bc}{\begin{Corrolary}}	\newcommand{\ec}{\end{Corrolary}}
\title{Holographic Gravitational Anomaly and Chiral Vortical Effect }

\author{Karl Landsteiner\thanks{karl.landsteiner@csic.es}, 
Eugenio Meg\'{\i}as\thanks{eugenio.megias@csic.es}, 
Luis Melgar\thanks{luis.melgar@estudiante.uam.es}, 
Francisco Pena-Benitez\thanks{fran.penna@uam.es}\\

Instituto de F\'{\i}sica Te\'orica UAM/CSIC, C/ Nicol\'as Cabrera 13-15,\\
Universidad Aut\'onoma de Madrid, Cantoblanco, 28049 Madrid, Spain}
\abstract{We analyze a holographic model with a pure gauge and a mixed
  gauge-gravitational Chern-Simons term in the action. These are the holographic implementations of the usual chiral and the mixed
  gauge-gravitational anomalies in four dimensional field theories with
  chiral fermions.  We discuss the holographic renormalization and
  show that the gauge-gravitational Chern-Simons term does not induce
  new divergences. In order to cancel contributions from the
  extrinsic curvature at a boundary at finite distance a new type of
  counterterm has to be added however. This counterterm can also serve to make
  the Dirichlet problem well defined in case the gauge field strength
  vanishes on the boundary. A charged asymptotically AdS black hole is
  a solution to the theory and as an application we compute the 
  chiral magnetic and chiral vortical conductivities via Kubo
  formulas. We find that the characteristic term proportional to $T^2$
  is present also at strong coupling and that its numerical value is not
  renormalized compared to the weak coupling result.  }

\preprint{IFT-UAM/CSIC-11-45}
\keywords{Gauge-gravity correspondence, Anomalies, Chern-Simons terms,
Transport theory}

\begin{document}

\input{intro}

\input{holoano}

\input{renorma}

\input{kubo}

\input{discussion}

\appendix

\input{eqmot}
\input{app_renorma}

\input{equations_shear}
\hfill\newpage
\input{sol_w_0}\hfill

\begin{acknowledgments}
We thank D. Grumiller for Email conversation.  This work has been
supported by Plan Nacional de Altas Energ\'{\i}as FPA2009-07908,
Spanish MICINN Consolider-Ingenio 2010 Program CPAN (CSD2007-00042),
Comunidad de Madrid HEP-HACOS S2009/ESP-1473.  L.M. has been supported
by fellowhship BES-2010-041571. F.P. has been supported by fellowship CPI Comunidad de Madrid.
\end{acknowledgments}

\bibliographystyle{JHEP}
\bibliography{HoloGravAnomal}

\end{document}

%% file: intro.tex
\section{Introduction}

Anomalies belong to the most interesting and most subtle properties of
relativistic quantum field theories.  They are responsible for the
breakdown of a classical symmetry due to quantum effects. The
Adler-Bardeen non-renormalization theorem guarantees that this
breakdown is saturated at the one-loop level.  Therefore the presence
of anomalies can be determined through simple algebraic criteria on
the representations under which the chiral fermions of a particular
theory transform.  In vacuum the anomaly appears as the
non-conservation of a classically conserved current in a triangle
diagram with two additional currents. In four dimension two types of
anomalies can be distinguished according to whether only spin one
currents appear in the triangle \cite{Adler:1969gk,Bell:1969ts} or if also the
energy-momentum tensor participates \cite{Delbourgo:1972xb,Eguchi:1976db}. We will
call the first type of anomalies simply chiral anomalies and the
second type gravitational anomalies. To be precise, in four dimension
we should actually talk of mixed gauge-gravitational anomalies since
triangle diagrams with only energy-momentum insertions are perfectly
conserved (see
e.g. \cite{AlvarezGaume:1983ig}). In a
basis of only right-handed fermions transforming under a symmetry generated
by $T_A$ the presence of chiral anomalies
is detected by the non-vanishing of $d_{ABC} = \frac 1
2\mathrm{Tr}(T_A\{T_B,T_C\})$ whereas the presence of a gravitational
anomaly is detected by the non-vanishing of $b_A= \mathrm{Tr}(T_A)$.

It recently has been emphasized how at finite temperature and density
anomalies give rise to new non-dissipative transport phenomena in the
hydrodynamics of charged relativistic fluids 
\cite{Fukushima:2008xe,Erdmenger:2008rm,Banerjee:2008th,Son:2009tf}. More precisely
magnetic fields and vortices in the fluid induce currents via the
so-called chiral magnetic and chiral vortical conductivities. Although
there have been many early precursors that found manifestations of
this phenomena in the physics of neutrinos
\cite{Vilenkin:1995um,Vilenkin:1980ft,Vilenkin:1980zv,Vilenkin:1978hb},
the early universe \cite{Giovannini:1997eg}, condensed matter systems
\cite{Alekseev:1998ds}, the recent surge of interest is clearly
related to the physics of the quark gluon plasma.  It has been
suggested that the observed charge separation in heavy ion collisions
is related to a particular manifestation of these anomalous transport
phenomena, the chiral magnetic effect \cite{Kharzeev:2007jp}. The
latter describes how a usual (i.e. electro-magnetic) B-field induces
via the axial anomaly an electric current parallel to the magnetic
field.  The first application of holography to the anomalous
hydrodynamics is \cite{Newman:2005hd} where the anomalous transport
effects due to R-charge magnetic fields have been examined.  Later
studies showed that there is also a related vortical effect
\cite{Erdmenger:2008rm,Banerjee:2008th}, i.e. a vortex in the fluid
induces a current parallel to the axial vorticity vector $\omega^\mu =
\epsilon^{\mu\nu\rho\lambda} u_\nu \partial_\rho u_\lambda$, and
related effects of the presence of angular momentum had been discussed
before in a purely field theoretical setup in \cite{Vilenkin:1980zv}
and \cite{Kharzeev:2007tn}.  Studies of the chiral magnetic effect
using holography have appeared in
\cite{Yee:2009vw,Torabian:2009qk,Rebhan:2009vc,Brits:2010pw,Gynther:2010ed,Gorsky:2010xu,Kalaydzhyan:2011vx,Hoyos:2011us,Eling:2010hu}
and using lattice field theory in
\cite{Buividovich:2009wi,Abramczyk:2009gb,Yamamoto:2011gk}.  A related
effect is the so called chiral separation effect that induces an axial
current in a magnetic field \cite{Newman:2005as}. Chiral magnetic
waves have been shown to arise through the interplay of chiral
magnetic effect and chiral separation effect in
\cite{Kharzeev:2010gd}.  Experimental signatures of anomalous
transport beyond the charge separation effect have been proposed in
\cite{Kharzeev:2010gr,
  KerenZur:2010zw,Asakawa:2010bu,Ajitanand:2010rc}. The experimental
status of the observed charge separation in heavy ion collision is
discussed in \cite{abelev:2009uh,abelev:2009txa}.

A first principals approach to transport theory is via Kubo formulas. In general the reaction of a system to external perturbations can be studied via linear response theory. The basic objects of linear response theory are the retarded Green functions. Hydrodynamic transport coefficients can be extracted from the long-wavelength and low-frequency limits of the retarded Green functions. 
A typical example is the Kubo formula for the shear viscosity
\begin{equation}\label{eq:KuboShear}
\eta = \lim_{\omega\rightarrow 0} \frac{i}{\omega} \left\langle T_{xy} T_{xy} \right\rangle (\omega,\vec{p}=0)\,.
\end{equation}
Electric or thermal conductivities can be calculated in a similar fashion. 

Also the transport coefficients related to the presence of anomalies can be computed via Kubo formulas. For the chiral magnetic effect this has been done in \cite{Kharzeev:2009pj} and has been further studied in \cite{Hou:2011ze}.
The Kubo formula for the chiral vortical conductivity was derived in \cite{Amado:2011zx}. These formulas can easily
be generalized to the case of a general non-abelian symmetry group, generated by matrices $T_A$. 
For the chiral magnetic conductivities and chiral vortical conductivities they are
\begin{eqnarray}\label{eq:KuboCMC}
 \sigma_{AB}^{\cB} &=& \lim_{p_n\rightarrow 0}\frac{i}{2 p_c} \sum_{a,b} \epsilon_{abc} \left\langle J_A^a J_B^b \right\rangle (\omega=0,\vec{p})\,,\\
\label{eq:KuboCVC}
 \sigma_{A}^{\cV} &=& \lim_{p_c\rightarrow 0}\frac{i}{2 p_c} \sum_{a,b} \epsilon_{abc} \left\langle J_A^a T^{0b}\right\rangle(\omega=0,\vec{p})\,,
\end{eqnarray}
with $a,b,c=x,y,z$.

The properties of these Kubo formulas for the anomalous conductivities are rather different from the ones
for the dissipative conductivities. Whereas the Kubo formulas for the usual transport coefficients are given by the derivatives with respect to the frequency at zero momentum, the ones for the anomaly related transport coefficients 
are given by derivatives in momentum space at zero frequency. For this reason the anomalous conductivities are
contained in the hermitian part of the correlators. In contrast the dissipative transport coefficients are contained
in the spectral densities, i.e. the anti-hermitian parts of the correlators. In the presence of external sources
$f_I(\omega)$ that couple to operators $\cO^I$ the rate of dissipation is describe in terms of the spectral density 
$\rho^{IJ}$, i.e. the anti-hermitian part of the retarded correlator
$\langle\cO^I \cO^J \rangle(\omega, \vec{p})$, by \cite{LeBellac:1991cq}
\begin{equation}
\frac{d W}{dt} = \frac 1 2 \omega f_I(-\omega) \rho^{IJ}(\omega) f_J(\omega)\,.
\end{equation}
This shows that the anomaly related currents due to the conductivities (\ref{eq:KuboCMC}) and (\ref{eq:KuboCVC}) do
no work on the system and are therefore examples of dissipationless transport. Let us also note that
dissipative transport breaks time reversal invariance $\cT$ whereas anomaly induced
dissipationless transport preserves $\cT$\footnote{This point has recently also been emphasized in \cite{Kharzeev:2011ds}.}.

In \cite{Landsteiner:2011cp} these general Kubo formulas were
evaluated for a theory of free chiral fermions. The results showed a somewhat surprising appearance of the anomaly coefficient $b_A$ for the gravitational anomaly. More precisely the chiral vortical conductivity for the symmetry generated by $T_A$ was found to have two contributions, one depending only on the chemical potentials and proportional to the axial anomaly coefficient $d_{ABC}$ and a second one with a characteristic $T^2$ temperature dependence proportional to the gravitational anomaly coefficient $b_A$.
At weak coupling the anomalous magnetic and vortical conductivities were found to be
\begin{eqnarray}
\sigma^B_{AB} &=& \frac{d_{ABC}}{4 \pi^2} \mu^C\,\label{eq:sigmaB} \,,\\
\sigma^V_A &=& \frac{d_{ABC}}{8 \pi^2} \mu^B \mu^C + \frac{b_A}{24} T^2\,.\label{eq:sigmaV}
\end{eqnarray}

This characteristic $T^2$ behavior had appeared already previously in neutrino physics \cite{Vilenkin:1978hb,Vilenkin:1980zv,Vilenkin:1980ft,Vilenkin:1995um}. It furthermore shows up
via undetermined integration constants in effective field theory inspired approaches to hydrodynamics. 
A purely hydrodynamic approach to anomalous transport has been initiated in \cite{Son:2009tf} (see
also \cite{Sadofyev:2010pr}). There it was
shown that the anomalous transport coefficient can be fixed by construction of an appropriate entropy current whose
divergence is positive definite.
Later in \cite{Neiman:2010zi} it was shown that there are additional integration constants proportional to
$T^2$ and $T^3$. CPT invariance forbids
however the $T^3$ terms \cite{Bhattacharya:2011tr} so that at least in systems that can be described by local quantum field theories these terms are absent. Recently these studies have been extended to
superfluids \cite{Lin:2011mr,Bhattacharya:2011tr,Neiman:2011mj}, to second order hydrodynamics 
\cite{Kharzeev:2011ds} and to higher dimension \cite{Loganayagam:2011mu} where similar undetermined
integration constants were found.

The usage of Kubo formulas has here a clear advantage, it fixes
all integration constants automatically. In this way it was possible in \cite{Landsteiner:2011cp} to show that the coefficient in front
of the $T^2$ term in the chiral vortical conductivity is essentially given by the gravitational anomaly coefficient $b_A$. The disadvantage of Kubo formulas is of course that we have to calculate the potentially complicated correlations functions of a quantum field theory. They are easy to evaluate only in certain limits, such as the weak coupling limit considered in \cite{Landsteiner:2011cp}. In principle the results obtained in this limit can suffer renormalization due
to the model dependent interactions. 

The gauge-gravity correspondence \cite{Maldacena:1997re,Gubser:1998bc,Witten:1998qj,Aharony:1999ti} makes also the strong coupling limit easily accessible. Strongly coupled non-abelian
gauge theories can be described via their gravity duals, more precisely in the large $N$ and infinite 't-Hooft coupling limit, $g^2 N\rightarrow\infty$, allows a weakly coupled gravitational description. 
The drawback is that only some special and supersymmetric gauge theories, such as the maximally supersymmetric $N=4$ Yang-Mills theory have well understood gravity duals. 

We would like to understand the effects anomalies have on the
transport properties of relativistic fluids.  Anomalies are very
robust features of quantum field theories and do not depend on the
details of the interactions.  Therefore a rather general model that
implements the correct anomaly structure in the gauge-gravity setup is
sufficient for our purpose even without specifying in detail to which
gauge theory it corresponds to.  Our approach will therefore be a
``bottom up'' approach in which we simply add appropriate Chern-Simons
terms that reproduce the relevant anomalies to the Einstein-Maxwell
theory in five dimensions with negative cosmological constant.~\footnote{Very successful holographic bottom up approaches to QCD have been studied recently, either to describe non-perturbative phenomenology in the vacuum, see e.g.~\cite{Andreev:2006ct,Galow:2009kw}, or the strongly coupled plasma~\cite{Gursoy:2008za,Megias:2010ku,Veschgini:2010ws}.}

In this paper we will introduce a model that allows for a holographic
implementation of the mixed gauge-gravitational anomaly via a mixed
gauge-gravitational Chern-Simons term of the form
\begin{equation}
S_{CS} = \int d^5x \, \sqrt{-g} \epsilon^{MNPQR} A_M R^A\,_{BNP} R^B\,_{AQR} \,.
\end{equation}

Gravity in four dimensions  augmented by a similar term with a scalar field instead of a vector field has attracted
much interest recently \cite{Jackiw:2003pm} (see also the review \cite{Alexander:2009tp}). A four dimensional
holographic model with such a term has been shown to give rise to Hall viscosity in \cite{Saremi:2011ab}. The 
quasinormal modes of this four dimensional model have been studied in \cite{Delsate:2011qp}.

In section~\ref{sec:holo_model} we will define the Lagrangian of our
model, derive its equations of motion and study how the gravitational
anomaly arises. We will find it necessary to add a particular boundary
counterterm that cancels dependences on the extrinsic curvature of
the gauge variation of the action. In section~\ref{sec:holo_renorm} we
will study the holographic renormalization assuming the existence of
an asymptotically AdS solution and show that the gravitational
Chern-Simons terms does not introduce new divergencies.

In section~\ref{sec:kubo} we will compute the equations of motions for
metric and gauge perturbations in the shear sector.  At zero frequency
and to lowest order in the momentum these equations can be solved
analytically. This allows us to compute the anomaly related
conductivities in our model. We find that the contribution due to the
gravitational anomaly is not renormalized compared to the weak
coupling result. As is well-known, in a charged fluid a current
necessarily induces also an energy flux. Kubo formulas for the energy
flux induced by magnetic fields and vortices related to
energy-momentum correlators. We also evaluate these and find that no
terms proportional to $T^3$ appear. This is of course consistent with
CPT conservation. We also show that the shear-viscosity to entropy 
ratio is unchanged. 

We conclude with a discussion of our results and an outlook towards
possible future directions in section~\ref{sec:discussion}.  Several
technical details of the calculations such as the Gauss-Codazzi form
of the equations of motion, the details of the holographic
renormalization, and the equations of motion for the shear sector and
their solutions are collected in the appendices.

%% file: holoano.tex
\section{Holographic Model}
\label{sec:holo_model}

In this section we will define our model. 
We start by fixing our conventions.
We choose the five dimensional metric to be of signature $(-,+,+,+,+)$. The epsilon tensor has to be distinguished from the epsilon symbol.
The symbol is defined by $\epsilon(rtxyz) = +1$ whereas the tensor is defined by $\epsilon_{ABCDE}=\sqrt{-g}\,\epsilon(ABCDE)$. Five dimensional indices
are denoted with upper case latin letters. We define an outward pointing normal vector $n_A \propto g^{AB} \frac{\partial r}{\partial x^B}$ to the
holographic boundary of an asymptotically AdS space with unit norm $n_A n^A =1$ so that the induced metric takes the form 
\begin{equation}\label{eq:inducemetric}
 h_{AB} = g_{AB}- n_A n_B\,.
\end{equation}

In general a foliation with timelike surfaces defined through $r(x) = \mathrm{const}$ can be written as
\begin{equation}
 ds^2 = (N^2 + N_A N^A) dr^2 + 2N_A dx^A dr+ h_{AB}dx^A dx^B\,.
\end{equation}

The Christoffel symbols,  Riemann tensor and extrinsic curvature are given by
\begin{eqnarray}
 \Gamma^M_{NP} &=& \frac 1 2 g^{MK}\left(  \partial_N g_{KP} +  \partial_P g_{KM} - \partial_K g_{NP}  \right),\\ 
 R^M\,_{NPQ} &=& \partial_P \Gamma^M_{NQ} - \partial_Q \Gamma^M_{NP} + \Gamma^M_{PK} \Gamma^K_{NQ} -   \Gamma^M_{QK} \Gamma^K_{NP} ,\\
 K_{AV} &=&   h_A^C \nabla_C n_V =  \frac 1 2 {\pounds}_n h_{AB} \,,
\end{eqnarray}
where $\pounds_n$ denotes the Lie derivative in direction of $n_A$  .

Finally we can define our model. The action is given by \bea\nonumber
S &=& \frac{1}{16\pi G} \int d^5x \sqrt{-g} \left[ R + 2 \Lambda -
  \frac 1 4 F_{MN} F^{MN} \right.\\ &&\left.+ \epsilon^{MNPQR} A_M
  \left( \frac\kappa 3 F_{NP} F_{QR} + \lambda R^A\,_{BNP} R^B\,_{AQR}
  \right) \right] + S_{GH} + S_{CSK} \,,\\ S_{GH} &=& \frac{1}{8\pi G}
\int_\partial d^4x \sqrt{-h} \, K \,,\\ S_{CSK} &=& - \frac{1}{2\pi G}
\int_\partial d^4x \sqrt{-h} \, \lambda n_M \epsilon^{MNPQR} A_N
K_{PL} D_Q K_R^L \,, \eea where $S_{GH}$ is the usual Gibbons-Hawking
boundary term and $D_A=h_A^B\nabla_B$ is the covariant derivative on
the four dimensional boundary. The second boundary term $S_{CSK}$ is
needed if we want the model to reproduce the gravitational anomaly at
general hypersurface.  To study the behavior of our model under the
relevant gauge and diffeomorphism gauge symmetries we note that the
action is diffeomorphism invariant. The Chern Simons terms are well
formed volume forms and as such are diffeomorphism
invariant. They do depend however explicitly on the gauge connection
$A_M$.  Under gauge transformations $\delta A_M = \nabla_M \xi$ they
are therefore invariant only up to a boundary term. We have
\begin{eqnarray}
 \delta S &=& \frac{1}{16\pi G} \int_\partial d^4x \sqrt{-h} \,\xi  \epsilon^{MNPQR} \left( \frac{\kappa}{3}n_M F_{NP}F_{QR} +
\lambda n_MR^A\,_{BNP} R^{B}\,_{AQR}\right) -\,\nonumber\\
& &- \frac{\lambda}{4\pi G} \int_\partial d^4x \sqrt{-h} \,n_M \epsilon^{MNPQR}D_N \xi K_{PL}D_Q K^L_R \,.
\end{eqnarray}
This is easiest evaluated in Gaussian normal coordinates (see next
section) where the metric takes the form $ds^2 = dr^2 + \gamma_{ij}
dx^i dx^j$. All the terms depending on the extrinsic curvature cancel
thanks to the contributions from $S_{CSK}$! The gauge variation of the
action depends only on the intrinsic four dimensional curvature of the
boundary and is given by
\begin{equation}
 \delta S = \frac{1}{16 \pi G} \int_\partial d^4 x \sqrt{-h} \epsilon^{mnkl}\left( \frac{\kappa}{3} \hat{F}_{mn} \hat{F}_{kl} + \lambda \hat R^{i}\,_{jmn} \hat R^{j}\,_{ikl}\right) \,.
\end{equation}
This has to be interpreted as the anomalous variation of the effective quantum action of the dual field theory.
The anomaly is therefore in the form of the consistent anomaly. Since we are dealing only with a single $U(1)$ 
symmetry the (gauge) anomaly is automatically expressed in terms of the field strength. We could also express
the anomaly in terms of an anomalous current conservation equation. One has to be however careful about the definition
of the current since it is always possible to add a Chern-Simons current and redefine $J^m \rightarrow J^m + c \epsilon^{mnkl}A_n F_{kl}$. This redefined current can not be expressed as the variation of a local functional of
the fields with respect to the gauge field. In particular the so-called covariant form of the anomaly differs precisely
in such a redefinition of the current\footnote{Note that the effective field thoery hydrodynamic approaches
following \cite{Son:2009tf} typically use the covariant form of the anomaly 
\cite{Neiman:2010zi}.}. 
A good general reference for anomalies is Bertlmann's book \cite{Bertlmann:1996xk}
where the consistent form of the anomaly for chiral fermions
transforming under a symmetry group generated by $T_A$ is quoted as
\begin{equation}\label{eq:consistent}
D_m J^m_A = \eta_H \frac{1}{24\pi^2} \epsilon^{ijkl}
\mathrm{Tr}\left[ T_A \partial_i \left( A_j \partial_k A_l + \frac 1 2
    A_j A_k A_l \right) \right]\,,
\end{equation}
with $\eta_H = \pm$ for $H\in\{R,L\}$ for right-handed and left-handed
fermions respectively.
We use this to fix $\kappa$ to the anomaly coefficient for a single
chiral fermion transforming under a $U(1)_L$ symmetry. To do so we simply
set $T_A=1$ in (\ref{eq:consistent}) which fixes the anomaly
coefficient $d_{ABC} = \frac 1 2 \mathrm{Tr}(T_A\{T_B,T_C\}) = 1$ and therefore
\begin{equation}
-\frac{\kappa}{48 \pi G} = \frac{1}{96 \pi^2}\,. 
\end{equation}
Similarly we can fix $\lambda$ by matching to the gravitational
anomaly of a single left-handed fermion
\begin{equation}
D_m J^m = \frac{1}{768\pi^2} \epsilon^{ijkl} \hat{R}^m\,_{nij} \hat{R}^n\,_{mkl} \,,
\end{equation}
and find
\begin{equation}
-\frac{\lambda}{16\pi G} = \frac{1}{768 \pi^2} \,.
\end{equation}
As a side remark we note that the gravitational anomaly could in principle also be shifted into the diffeomorphism sector. This can be done
by adding an additional (Bardeen like) boundary counterterm to the action 
\begin{equation}\label{eq:gravBardeen}
 S_{ct} = \int d^4x \, \sqrt{-h} A_m I^m\,,
\end{equation}
with $I^m=\epsilon^{mnkl}( \hat\Gamma^p_{nq}\partial_k \hat\Gamma^q_{lp} + \frac 2 3 \hat\Gamma^o_{mp}\hat\Gamma^p_{kq}\hat\Gamma^q_{lo})$ fulfilling $D_m I^m = \frac 1 4 \epsilon^{ijkl} \hat{R}^m\,_{nij} \hat{R}^n\,_{mkl}$. Since this term depends explicitly on the four dimensional Christoffel connection it
breaks diffeomorphism invariance. 

The bulk equations of motion are
\begin{eqnarray}\label{eqgrav}
 G_{MN} - \Lambda g_{MN} &=& \frac 1 2 F_{ML} F_N\,^L - \frac 1 8 F^2 g_{MN} + 2 \lambda \epsilon_{LPQR(M} \nabla_B\left( F^{PL} R^B\,_{N)}\,^{QR} \right) \,, \label{eq:Gbulk}\\\label{eqgauge}
\nabla_NF^{NM} &=& - \epsilon^{MNPQR} \left( \kappa F_{NP} F_{QR} + \lambda  R^A\,_{BNP} R^B\,_{AQR}\right) \,,  \label{eq:Abulk}
\end{eqnarray}
and they are gauge and diffeomorphism covariant. We note that keeping all
boundary terms in the variations that lead to the bulk equations of
motion we end up with boundary terms that contain derivatives of the
metric variation normal to the boundary. We will discuss this issue in
more detail in the next section where we write down the Gauss-Codazzi
decomposition of the action.

%% file: renorma.tex
\section{Holographic Renormalization}
\label{sec:holo_renorm}

In order to go through the steps of the holographic renormalization
program within the Hamiltonian approach \cite{Martelli:2002sp,Papadimitriou:2004ap},
first of all we establish some notations.  Without loss of generality
we choose a gauge with vanishing shift vector $N_A=0$, lapse $N=1$ and
$A_r=0$. So we can use four dimensional (boundary) indices and denote
them by small latin letters. We therefore also write
$\epsilon(txyz)=+1$ and
$\epsilon_{ijkl}=\sqrt{-h}\,\epsilon(ijkl)$. In this gauge the bulk
metric can be written as
\begin{equation}
 ds^2 = dr^2 + \gamma_{ij} dx^i dx^j \,.
\end{equation}
The non vanishing Christoffel symbols are
\begin{eqnarray}
-\Gamma^r_{ij} &=&  K_{ij} = \frac 1 2 \dot{\gamma}_{ij} \, ,\\
\Gamma^i_{jr} &=& K^i_j \,,  
\end{eqnarray}
and $\hat\Gamma^i_{jk}$ are four dimensional Christoffel symbols computed with $\gamma_{ij}$. Dot denotes differentiation respect $r$. All other components of the extrinsic curvature vanish, i.e. $K_{rr}=K_{ri} =0$. Another useful table of formulas is
\begin{eqnarray}
\dot{\hat\Gamma}^l_{\,ki} \,\,\,&=& D_k K^l_i + D_i K^l_k - D^l K_{ki} \,, \\
R^r\,_{irj} &=& -\dot{K}_{ij} + K_{il}K^l_j \,,\\
R^k\,_{rjr} &=& -\dot{K}^k_{j} - K^k_{l}K^l_j \,,\\
R^r\,_{ijk} &=& D_k K_{ij} - D_j K_{ik} \,,\\
R^l\,_{kri} &=& D_k K^l_i- D^lK_{ik} \,,\\
R^i\,_{jkl} &=& \hat{R}^i\,_{jkl} - K^i_k K_{jl} +  K^i_l K_{jk}\,.
\end{eqnarray}
Note that indices are now raised and lowered with $\gamma_{ij}$, e.g. $K=\gamma^{ij} K_{ij}$, and intrinsic four dimensional curvature quantities are 
denoted with a hat, so $\hat{R}^i\,_{jkl}$ is the intrinsic four dimensional Riemann tensor on the $r=\textrm{const}$ surface. Finally the Ricci scalar is
\begin{equation}
 R = \hat{R} - 2 \dot{K} - K^2 - K_{ij}K^{ij} \,.
\end{equation}

Now we can calculate the off shell action. It is useful to divide it up in three terms. The first one is the usual gravitational bulk and gauge terms with the usual Gibbons-Hawking term. After some computations  we get
\begin{eqnarray}
\label{eq:Sb1} 
S^0 &=& \frac{1}{16 \pi G} \int  d^5x\,\sqrt{-\gamma} \left[ \hat{R} + 2 \Lambda + K^2 -
K_{ij}K^{ij} - \frac 1 2 E_i E^i - \frac 1 4 \hat{F}_{ij}\hat{F}^{ij} \right] \, ,\\
\label{eq:Sb2}
S^{1}_{CS} &=& -\frac{\kappa}{12\pi G} \int d^5x\,\sqrt{-\gamma}  \epsilon^{ijkl} A_i E_j \hat{F}_{kl}  \,,\\
\label{eq:Sb3}
S^2_{CS} &=& -\frac{8 \lambda}{16\pi G} \int d^5x \sqrt{-\gamma}  \epsilon^{ijlk} \bigg[ 
A_i \hat{R}^n\,_{mkl} D_nK^m_j + E_i K_{jm}D_k K^m_l + \frac 1 2 \hat{F}_{ik} K_{jm}\dot{K}^m_l \bigg] \,. 
\end{eqnarray}
We have used implicitly here the gauge $A_r=0$ and denoted $\dot{A}_i= E_i$. The purely four dimensional field strength is denoted with a hat. 

Of particular concern is the last term in $S^2_{CS}$ which contains
explicitly the normal derivative of the extrinsic curvature $\dot
K_{ij}$. For this reason the field equations will be generically of third
order in $r$-derivatives and that means that we can not define a
well-posed Dirichlet problem by fixing the $\gamma_{ij}$ and $K_{ij}$ alone but
generically we would need to fix also $\dot K_{ij}$.
Having applications to holography in mind we can however impose the
boundary condition that the metric has an asymptotically AdS expansion
of the form
\begin{equation}
\gamma_{ij} = \e^{2 r} \left( g^{(0)}_{ij} + e^{-2r} g^{(2)}_{ij} + e^{-4r} (g^{(4)}_{ij}
  + 2r \tilde{g}^{(4)}_{ij} )+ \cdots\right) \,.
\end{equation}
Using the on-shell expansion of $K_{ij}$ obtained in the appendix \ref{sec:app_holo_renorm}
we can show that the 
last term in the action does not contribute in the limit
$r\rightarrow\infty$. Therefore the boundary action depends only on the
boundary metric $\gamma_{ij}$ but not on the derivative $\dot
\gamma_{ij}$.  This is important because otherwise the dual theory
would have additional operators that are sourced by the
derivative. Similar issues have arisen before in the holographic
theory of purely gravitational anomalies of two dimensional field
theories \cite{Saremi:2011ab,Yee:2011yn,Kraus:2005zm}.  Alternatively
one could restrict the field space to configurations with vanishing
gauge field strength on the boundary. Then the last term in $S^2_{CS}$
is absent. We note that the simple form of the higher derivative terms
arises only if we include $S_{CSK}$ in the action. An analogous term in
four dimensional Chern-Simons gravity has been considered before in
\cite{Grumiller:2008ie}.

The renormalization procedure follows from an expansion of the four dimensional quantities in eigenfunctions of the dilatation operator
\begin{equation}
\delta_D = 2 \int d^4x \gamma_{ij} \frac{\delta}{\delta \gamma_{ij}} \,. 
\end{equation}
We explain in much details the renormalization in appendix~\ref{sec:app_holo_renorm}. The result one gets for the counterterm coming from the regularization of the boundary action is
\begin{eqnarray}
S_{ct} &=& - \frac{(d-1)}{8\pi G} \int_\partial d^4x \sqrt{-\gamma} \bigg[
1 + \frac{1}{(d-2)}P  \nonumber  \\
&&\qquad\qquad- \frac{1}{4(d-1)} \left( P^i_j P^j_i - P^2 -  \frac{1}{4} \hat{F}_{(0)}\,_{ij} \hat{F}_{(0)}\,^{ij} \right)\log e^{-2r} \bigg] \,, \label{eq:Sct}
\end{eqnarray}
where
\begin{equation}
P = \frac{\hat{R}}{2(d-1)} \,, \qquad  P^i_j = \frac{1}{(d-2)} \left[ \hat{R}^i_j - P \delta^i_j \right] \,. \label{eq:PPij} 
\end{equation}
As a remarkable fact there is no contribution in the counterterm
coming from the gauge-gravitational Chern-Simons term. This has also
been derived in \cite{Clark:2010fs} in a similar model that does
however not contain $S_{CSK}$.

%% file: kubo.tex
\section{Kubo formulas, anomalies and chiral vortical conductivity}
\label{sec:kubo}
We are now going to evaluate the Kubo formulas for anomalous transport 
in our holographic model. First we note that
in a charged fluid a charge current is always accompanied by an energy
current through $\delta T^{0a} = \mu \delta J^a$. Therefore charge
transport is always accompanied by energy transport. Kubo formulas for
the energy transport coefficients can easily be obtained as well.  In
\cite{Amado:2011zx} it was shown that the chiral vortical conductivity
for charge and energy transport can be obtained from the retarded
Green functions
\begin{eqnarray}
\label{eq:sigmaV1}\sigma_V  &=& \lim_{k_c\rightarrow 0} \frac{i}{2k_c} \sum_{a,b}\epsilon_{abc}
\langle J^a T^{0b} \rangle|_{\omega=0} \, ,\\
 \label{eq:sigmaVe}  \sigma_V^\epsilon  &=& \lim_{k_c\rightarrow 0} \frac{i}{2k_c}\sum_{a,b}\epsilon_{abc}
\langle T^{0a} T^{0b} \rangle|_{\omega=0}  \,,
 \end{eqnarray}
where $J^i$ is the (anomalous) current and $T^{ij}$ is the energy
momentum tensor,  $\sigma_V$ the chiral vortical conductivity and $\sigma_V^\epsilon$ the vortical  conductivity of energy current. The chiral magnetic conductivity $\sigma_B$ and the magnetic conductivity for energy current $\sigma_B^\epsilon$ are given by
\bea
\label{eq:sigmaB1}
\sigma_B&=& \lim_{k_c\rightarrow 0} \frac{i}{2k_c} \sum_{a,b}\epsilon_{abc}
\langle J^a J^b \rangle|_{\omega=0,A_0=0} \,,\\
\label{eq:sigmaBe}
\sigma_B^\epsilon&=& \lim_{k_c\rightarrow 0}\frac{i}{2k_c} \sum_{a,b}\epsilon_{abc}
\langle T^{0a} J^b \rangle|_{\omega=0} \,.
\eea
As explained in \cite{Gynther:2010ed,Amado:2011zx} we also have to set the background value of the temporal component of the gauge field to zero. 
Hydrodynamic constitutive relations depend however on a particular definition of the fluid velocity.
In the case of the anomalous conductivities this frame dependence has been addressed in \cite{Amado:2011zx} where
it was shown how the Landau frame conductivities used by Son \& Surowka \cite{Son:2009tf} can be obtained
from a combination of the charge and energy transport coefficient. This combination emerges because of the change of coordinates from the laboratory rest frame to a local comoving frame on a element of fluid in which there is no energy flux. Applying this change of frame we arrive to the transport coefficients in Landau frame
\begin{eqnarray}
 \label{eq:cmc}\xi_{ B}  &=& \lim_{k_c\rightarrow 0} \frac{i}{2 k_c} \sum_{a,b} \epsilon_{abc} \left(\left. \left\langle J^{a} J^{b} \right\rangle - 
\frac{n}{\epsilon+P}\left\langle T^{0a} J^{b} \right\rangle \right)\right|_{\omega=0, A_0=0}\,,\\
\label{eq:cvc}\xi_{V} &=& \lim_{k_c\rightarrow 0} \frac{i}{2 k_c} \sum_{a,b} \epsilon_{abc} \left.\left( \left\langle J^a T^{0b} \right\rangle - 
\frac{n}{\epsilon+P}\left\langle T^{0a} T^{0b} \right\rangle \right) \right|_{\omega=0}\,.
\end{eqnarray}
The frame dependence has also recently been discussed in \cite{Loganayagam:2011mu}. The relevant parts of the hydrodynamic constitutive relations are
\begin{eqnarray}
 \delta T^{mn} &=& \sigma_B^\epsilon (u^m B^n + u^n B^m) + \sigma_V^\epsilon (u^m \omega^n + u^n \omega^m)\,,\\
\delta J^{m} &=& \sigma_B  B^m + \sigma_V \omega^m \,,
\end{eqnarray}
whereas in Landau frame demanding $u_m \delta T^{mn}=0$ we have no contribution to the energy momentum tensor but instead
\begin{equation}
 \delta J^{m} = \xi_B  B^m + \xi_V \omega^m \,,
\end{equation}
where $B^m=\frac 1 2 \epsilon^{mnkl}u_n F_{kl}$.

The AdS/CFT dictionary  tells us how to compute the retarded propagators \cite{Son:2002sd,Herzog:2002pc}. Since we are interested in the linear response limit, we split the metric and gauge field into a background part and a linear perturbation,
\bea
g_{MN} &=& g^{(0)}_{MN} + \epsilon \, h_{MN} \,,\\ 
A_{M} &=& A^{(0)}_{M} + \epsilon \, a_{M} \, .
\eea
Inserting these fluctuations-background fields in the action and expanding up to second order in $\epsilon$ we can read the second order action which is needed to get the desired propagators~\cite{Kaminski:2009dh}.  If we construct a vector $\Phi^I$ with the components of $a_M$ and $h_{MN}$ and Fourier transforming it
\be 
\Phi^I(r,x^{\mu})=\int \frac{\dd^d k}{(2\pi)^d} \Phi^I_k (r) \e^{-i \omega t+i \vec{k}\vec{x}} \,,
\ee
it is possible to write the complete second order action on-shell as a boundary term
\be \label{eq:2ndor}
\delta S^{(2)}_{ren}=\int \frac{\dd^d k}{(2\pi)^d} \lbrace \Phi^I_{-k} \cA_{IJ} \Phi '^J_k + \Phi^I_{-k}  \cB_{IJ} \Phi^J_k \rbrace\Big{|}_{r\to\infty}\,,
\ee
where derivatives are taken with respect to the radial coordinate.

Now we can compute the holographic response functions
from~(\ref{eq:2ndor}) by applying the prescription of
\cite{Son:2002sd,Herzog:2002pc,Amado:2009ts,Kaminski:2009dh}.  
For a coupled system the holographic computation of the correlators consists
in finding a
maximal set of linearly independent solutions that satisfy infalling
boundary conditions on the horizon and that source a single operator
at the AdS boundary. To do so we can construct a matrix of solutions
$F^I\,_J (k,r)$ such that each of its columns corresponds to one of
the independent solutions and normalize it to the unit matrix at the
boundary. Therefore, given a set of boundary values for the
perturbations, $\varphi^I_k$, the bulk solutions are 
\be\label{eq:f}
\Phi^I_k (r) = F^I\,_J (k,r)\, \varphi^J_k\,.  
\ee 
Finally using this
decomposition we obtain the matrix of retarded Green functions
\be\label{eq:GR} 
G_{IJ}(k)= -2 \lim_{r\to\infty} \left(\cA_{IM}
(F^M\,_J (k,r))' +\cB_{IJ}\right)\, .  
\ee

The system of equations (\ref{eqgrav})-(\ref{eqgauge}) admit the
following exact background AdS Reissner-Nordstr\"om black-brane
solution \bea \dd s^2&=& \frac{\bar r^2}{L^2}\left(-f(\bar r) \dd t^2
+\dd \vec{x}^2\right)+\frac{L^2}{\bar r^2 f(\bar r)} \dd \bar
r^2\,,\nn\\ A^{(0)}&=&\phi(\bar r)\dd t = \left(\beta-\frac{\mu \,\bar
  r_{{\rm H}}^2}{\bar r^2}\right)\dd t\,, \eea where the horizon of
the black hole is located at $\bar r=\bar r_{\rm H}$ and the
blackening factor of the metric is \be f(\bar r)=1-\frac{M L^2}{\bar
  r^4}+\frac{Q^2 L^2}{\bar r^6}\,.  \ee The parameters $M$ and $Q$ of
the RN black hole are related to the chemical potential $\mu$ and the
horizon $\bar r_H$ by \be M=\frac{\bar r_{\rm
    H}^4}{L^2}+\frac{Q^2}{\bar r_{\rm H}^2}\quad,\quad Q=\frac{\mu\,
  \bar r_{\rm H}^2}{\sqrt{3}}\,.  \ee The Hawking temperature is given
in terms of these black hole parameters as \be T=\frac{\bar r_{\rm
    H}^2}{4\pi\, L^2} f(\bar r_{\rm H})' = \frac{ \left(2\, \bar
  r_{\rm H}^2\, M - 3\, Q^2 \right)}{2 \pi \,\bar r_{\rm H}^5} \,.
\ee The pressure of the gauge theory is $P=\frac{M}{16\pi GL^3}$ and
its energy density is $\epsilon=3P$ due to the underlying conformal
symmetry .

Without loss of generality we consider perturbations of momentum $k$ in the $y$-direction at zero frequency. To study the effect of anomalies  we just  turned on the shear sector (transverse momentum fluctuations) $a_\alpha$ and $h^\alpha_{\,t}$, where $\alpha=x,z$.\footnote{Since we are in the zero frequency case the fields $h_y^\alpha$ completely decouple of the system and take a constant value, see appendix \ref{eq_shear}. } For convenience we redefine new parameters and radial coordinate
\be
\bar\lambda=\frac{4\mu \lambda L^3}{\bar r_H^2}\quad;\qquad \bar \kappa=\frac{4\mu \kappa L^3}{\bar r_H^2}\quad;\qquad a=\frac{\mu^2L^2}{3\bar r_{\rm H}^2}\quad;\qquad u=\frac{\bar r_H^2}{r^2}\,.
\ee
Now the horizon sits at $u=1$ and the AdS boundary at $u=0$. Finally we can write the system of differential equations for the shear sector, that consists on four second order equations. Since we are interested in computing correlators at hydrodynamics regime, we will solve the system up to first order in $k$.  The reduced system can be written as

\bea
\label{htx1} \nonumber 0&=&  h^{\alpha''}_t(u) - \frac{h^{\alpha'}_t(u)}{u} 
- 3a u B'_\alpha(u) +i\bar\lambda k \epsilon_{\alpha\beta}\left[\left(24a u^3-6(1-f(u))\right)\frac {B_\beta(u)}{u}\right.\\
    &&\left.+(9a u^3-6(1-f(u)))B'_\beta(u)+2 u (uh^{\beta'}_t ( u ) )' \right] \,, \\ 
\nonumber\label{bx1} 0&=& B_\alpha'' ( u )+\frac { f' ( u ) }{f(u)}B_\alpha'( u )-  \frac{ h^{\alpha'}_t ( u
 )}{f(u)} \\
 &&+ik\epsilon_{\alpha\beta}\left(\ \frac{3}{u f(u)} \bar\lambda \left( \frac{2}{a} (f(u)-1)+ 3u^3 \right) h_t^{\beta'}(u)+\bar \kappa \frac{B_\beta(u)}{f(u)} \right) \, ,
\eea
with the gauge field redefined as $B_\alpha=a_\alpha/\mu$. The complete system of equations depending on frequency and momentum is showed in appendix~\ref{eq_shear}. This system consists of six dynamical equations and two constraints.

In order to get solutions at first order in momentum we expand the
fields in the dimensionless momentum $p=k/4\pi T$ such as \bea
h^\alpha_t(u) &=& h^{(0),\alpha(u)}_t+p \,h^{(1)\alpha(u)}_t
\,,\\ B_\alpha(u) &=& B^{(0)}_\alpha(u)+p \,B^{(1)}_\alpha(u)\,.  \eea
The relevant physical boundary conditions on fields are:
$h^\alpha_t(0)=\tilde H^\alpha$, $B_\alpha(0)=\tilde B_\alpha$; where
the `tilde' parameters are the sources of the boundary operators.  The
second condition compatible with the ingoing one at the horizon is
regularity for the gauge field and vanishing for the metric
fluctuation \cite{Amado:2011zx}.

After solving the system perturbatively (see appendix~\ref{sol_w_0} for solutions), we can go back to the formula (\ref{eq:GR}) and compute the corresponding holographic Green functions. 
If we consider the vector of fields to be 
\be
\Phi_k^{\top} (u) = \Big{(} B_x(u) ,\, h^x_{\,t}(u)  ,\, B_z(u) ,\, h^z_{\,t}(u) \Big{)} \,,
\ee
the $\cA$ and $\cB$ matrices for that setup take the following form

\be
\cA=\frac{\bar r_{\rm H}^4}{16\pi G L^5} \,{\rm Diag}\left( -3a f,\, \frac{1}{u} ,\, -3a f,\, \frac{1}{u} \right) \,,
\ee

\bea
\hspace{-0.8cm}\cB_{AdS+\partial}=
\frac{\bar r_{\rm H}^4}{16\pi G L^5}
\left(
\begin{array}{cccc}
0 & -3a  & \frac{4 \kappa i k \mu^2 \phi L^5}{3 r_{\rm H}^4} & 0  \\
0 & -\frac{3}{u^2} &  0 & 0  \\
\frac{-4 \kappa i k \mu^2 \phi L^5}{3 r_{\rm H}^4} & 0 & 0 & -3a  \\
0 & 0 & 0 & -\frac{3}{u^2}  \\
\end{array}
\right)\,,
\eea

\bea
\cB_{CT}=
\frac{\bar r_{\rm H}^4}{16\pi G L^5}
\left(
\begin{array}{cccc}
\vspace{0.15cm}
0&  0 & 0 & 0 \\
\vspace{0.15cm}
0 & \frac{3}{u^2 \sqrt{f\,}} & 0 & 0  \\
\vspace{0.15cm}
0 & 0  & 0 & 0\\
\vspace{0.15cm}
0 & 0  & 0 & \frac{3}{u^2 \sqrt{f\,}}  \\
\end{array}
\right)\,,
\hspace{-0.8cm}\,
\eea
where $\cB=\cB_{AdS+\partial}+\cB_{CT}$. Notice that there is no contribution to the matrices coming from the Chern-Simons gravity part, the corresponding contributions vanish at the boundary. These matrices and the perturbative solutions are the ingredients to compute the matrix of propagators. Undoing the vector field redefinition introduced in (\ref{htx1}) and (\ref{bx1}) the non-vanishing retarded correlation functions at zero frequency are then
\bea
\label{eq:giti}G_{x,tx} &=& G_{z,tz} =\frac{\sqrt{3}\, Q}{4 \pi\, G\, L^3  }\,, \\
\label{eq:gxz}G_{x,z} &=& - G_{z,x} =\frac{i\, \sqrt{3}\, k\, Q\,  \kappa}{2 \pi\, G \, \bar r_{\rm{H}}^2}+\frac{i\, k\, \beta \, \kappa}{6\pi\, G  }\,,\\
G_{x,tz} &=&G_{tx,z} = -G_{z,tx}=-G_{tz,x}=\frac{3\, i\, k\, Q^2\,  \kappa }{4 \pi\,G\,  \bar r_{\rm{H}}^4}+\frac{2ik\lambda \pi T^2}{G} \,,\\
G_{tx,tx} &=& G_{tz,tz}=\frac{M}{16\pi\, G\, L^3 }\,,\\
\label{eq:gtiti}G_{tx,tz} &=& -G_{tz,tx}=+\frac{i\, \sqrt{3}\, k\, Q^3\, \kappa}{2\pi\, G\,\bar r_{\rm{H}}^6} + \frac{4\pi i\sqrt{3}  k Q T^2 \lambda}{G \,\bar r_{\rm H}^2}\,.
\eea

Using the Kubo formulas \erf{eq:cmc} and \erf{eq:cvc}  and setting the deformation parameter to zero we recover the conductivities  
\begin{eqnarray}
\label{eq:sigb}\sigma_B &=& -\frac{\sqrt{3}\,  Q \, \kappa}{2 \pi
  \,G\, \bar r_{\rm{H}}^2} =  \frac{ \mu}{4 \pi^2}\,,\\
\label{eq:sigv}\sigma_V &=&\sigma_B^\epsilon=-\frac{3\, Q^2\,  \kappa }{4 \pi\,G\, \bar  r_{\rm{H}}^4}-\frac{2\lambda \pi T^2}{G}= \frac{\mu^2}{8\pi^2} +\frac{T^2}{24} \,, \\
\label{eq:sigve}\sigma_V^\epsilon &=&-\frac{ \sqrt{3}\, Q^3\,
  \kappa}{2\pi\, G\, \bar r_{\rm{H}}^6} - \frac{4\pi \sqrt{3}   Q T^2
  \lambda}{G \,\bar r_{\rm H}^2} =\frac{\mu^3}{12 \pi^2}+\frac{\mu
  T^2}{12}\,. 
\end{eqnarray}
The first expression is in perfect agreement with the literature and
the second one shows the extra $T^2$ term predicted in
\cite{Landsteiner:2011cp}. In fact the numerical coefficients coincide
precisely with the ones obtained in weak coupling. This we take as a
strong hint that the anomalous conductivities are indeed completely
determined by the anomalies and are not renormalized beyond one loop.
We also point out that the $T^3$ term that appears as undetermined
integration constant in the hydrodynamic considerations in \cite{Neiman:2011mj}
should make its appearance in $\sigma_V^\epsilon$. We do not find 
any such term which is consistent with the argument that this term is
absent due to CPT invariance.

It is also interesting to write down the vortical and magnetic
conductivity as they appear  in the Landau frame,
\begin{eqnarray}
\xi_B &=& -\frac{\sqrt{3} Q ( ML^2+3 \bar r_{\rm H}^4)\kappa}{8\pi G M L^2 \bar r_{\rm H}^2}-\frac{\sqrt{3}Q\lambda\pi T^2}{GM} \nonumber \\
\label{eq:xib}&=& \frac{1}{4\pi^2}\, \left( \mu -  \frac{1}{2}\frac{n(\mu^2+\frac{T^2}{12})}{\epsilon+P}\right)\,,\\
\nonumber\xi_V &=&-\frac{3 Q^2  \kappa }{4\pi G  ML^2 }-\frac{2\pi\lambda T^2(\bar r_{\rm H}^6-2L^2Q^2)}{ G M L^2 \bar r_{\rm H}^2}\\
\label{eq:xiv}&=&  \frac{\mu^2}{8\pi^2}\, \left( 1 - \frac{2}{3}
  \frac{n \mu}{\epsilon+P}\right)+\frac{T^2}{24}\left(1-\frac{2 n\mu}{\epsilon+P}\right)\,.
\end{eqnarray}

Finally let us also note that the shear viscosity is not modified by the presence of the gravitational anomaly. We know that $\eta\propto\lim_{w\to 0}\frac{1}{w}<T^{xy}T^{xy}>_{k=0}$, so we should solve the system at $k=0$ for the fluctuations $h^i_y$  but the  anomalous coefficients always appear with a momentum $k$ as we can see in (\ref{eq_Hys}), therefore if we switch off the momentum, the system looks precisely as the theory without anomalies. In \cite{Bonora:2011mf} it has been shown that the black hole entropy doesn't depend on the extra mixed Chern-Simons term, therefore the shear viscosity entropy ratio remain the same in this model\footnote{For a four dimensional
holographic model with gravitational Chern-Simons term and a scalar field this has also been shown in \cite{Delsate:2011qp}.}.

%% file: discussion.tex
\section{Discussion and conclusion}
\label{sec:discussion}

We have defined a holographic bottom up model that implements the
mixed gauge-gra\-vi\-ta\-tio\-nal anomaly in four dimensional field theory via
a mixed gauge-gravitational Chern-Simons term. 
We have discussed its holographic renormalization and have shown that
the Chern-Simons term does not introduce new divergencies. As a first 
application of this theory we have computed the anomalous
magnetic and vortical conductivities from a charged black hole
background and have found the $T^2$ terms characteristic for the
contribution of the gravitational anomaly.

The most important result is certainly that the numerical values of
the conductivities coincide precisely with the ones obtained at weak
coupling in \cite{Landsteiner:2011cp}. This is a strong hint towards a
non-renormalization
theorem for the anomalous conductivities including the contributions
from the gravitational anomaly.

We have studied a holographic system with only one anomalous $U(1)$ symmetry.
It should however present no problem to generalize our calculation to the 
case with additional non-abelian symmetries and various types of mixed
anomalies, e.g. mimicking the usual interplay of axial and vector symmetries
where gauge Bardeen counterterms are necessary to implement the correct anomaly
structures in the currents \cite{Rebhan:2009vc,Gynther:2010ed}.

So far the contributions of the gravitational anomaly have shown up in
the calculations of Kubo formulas. It is however also possible to
calculate directly the constitutive relations of the hydrodynamics
of anomalous currents via the fluid/gravity approach of \cite{Bhattacharyya:2008jc}.
This very interesting question can be addressed within the model
presented in this paper \cite{workinprogress}.

%% file: eqmot.tex
\section{Codazzi form of Equations of Motion}
\label{appendixeq}
We project the equations of motion (\ref{eq:Gbulk}) and (\ref{eq:Abulk}) into the boundary surface and the orthogonal direction and rewrite them in terms of quantities at the regulated boundary. Doing so we get a set of two dynamical equations 
\begin{eqnarray}
\label{eq:Ab2}  0&=& \dot{E}^i + K E^i +  D_j \hat{F}^{ji} 
- 4\epsilon^{ijkl} \bigg( \kappa E_j \hat{F}_{kl} 
+ 4\lambda \dot{K}^s_j D_l K_{sk} + 2\lambda \hat{R}^s\,_{tkl} D_s K^t_j  \nonumber \\
&&\qquad\qquad\qquad\qquad\qquad\qquad\qquad\qquad + 4\lambda K_{ks} K^t_l D_t K^s_j  + 4\lambda K_{st} K^t_j D_l K^s_k \bigg) \,,
\end{eqnarray}
\begin{eqnarray}
\label{eq:Gb3} 0&=&\dot{K}^i_j + K K^i_j - \hat{R}^i_j 
+ \frac{1}{2}E^i E_j + \frac{1}{2} \hat{F}^{im}\hat{F}_{jm} - \frac{\delta^i_j}{(d-1)} \bigg( 2\Lambda + \frac{1}{2} E^m E_m + \frac{1}{4} \hat{F}^{lm}\hat{F}_{lm} \bigg) \nonumber \\
&&+2\lambda \bigg[
-2 \epsilon^{(iklm}\partial_r\left(\hat{F}_{kl}\dot{K}_{mj)}\right)
+2 \epsilon^{[iklm}\partial_r\left(\hat{F}_{kl}K_{ms}K^s_{j]}\right)
+2 \epsilon^{iklm}\hat{F}_{kl}K_{js}\left(\dot{K}^s_m + K^s_t K^t_m\right) \nonumber \\
&&\quad-\epsilon^{klmn}\hat{F}_{kl}\left( K^{(i}_s \hat{R}^s\,_{j)mn} 
+2 K^{(i}_m \dot{K}_{nj)} - 2 K^i_s K^s_m K_{nj}\right)  +4 \epsilon^{(iklm}\partial_r\left(E_k D_m K_{j)l} \right) \nonumber \\
&&\quad+2 \epsilon^{(iklm}D_s\left(\hat{F}_{kl}\left( D_ {j)} K^s_m - D^s K_{j)m} \right) \right) +4 \epsilon^{iklm} E_k K_{js} D_l K^s_m \nonumber \\
&&\quad-4 \epsilon^{klmn} E_k K^{(i}_l D_n K_{mj)} +2\epsilon^{(iklm} D_s\left( E_k (\hat{R}^s\,_{j)lm} - 2 K^s_l K_{j)m}) \right)  
\bigg] \,,
\end{eqnarray}
and three constraints
\begin{eqnarray}
 \label{eq:Gb1} 0&=&K^2 - K_{ij} K^{ij} - \hat{R} - 2 \Lambda - \frac{1}{2} E_i E^i + \frac{1}{4} \hat{F}_{ij}  \hat{F}^{ij} \nonumber \\
&&\qquad\qquad+ 8 \lambda \epsilon^{ijkl} \bigg( D_m ( \hat{F}_{ij} D_k K^m_l ) 
+  \hat{F}_{ij} K_{km} \dot{K}^m_l   + 2 E_i K_{jt} D_l K^t_k \bigg)  \,, \\
\label{eq:Gb2} 0&=&D_j K^{ji} - D^i K + \frac{1}{2} E_j \hat{F}^{ji} +  2 \lambda \epsilon^{klmi}  D_j\left[2E_k D_l K^j_m + \hat F_{kl}\left(\dot K^j_m + K^j_s K^s_m\right)\right] \nonumber\\
&&+\lambda\epsilon^{klmn}\left\{2\hat F_{kl}K^i_j D_mK^j_n + D_j\left[ F_{kl}(\hat R^{ij}\,_{nm}+2K^i_nK^j_m)\right] \right.\nonumber\\
&&\left. +2E_k K^j_m \hat R^i\,_{jnl} +2 \hat F_{kl} K^j_m(D^i K_{nj}-D_j K^i_n)
+2\partial_r(\hat F_{kl}D_nK^i_m)\right\}   \,,  \\
 \label{eq:Ab1}0&=&D_i E^i - \epsilon^{ijkl} \bigg( \kappa \hat{F}_{ij}\hat{F}_{kl} 
+ \lambda  \hat{R}^s\,_{tij} \hat{R}^t\,_{skl}
+ 4 \lambda K_{is} K^t_j\hat{R}^s\,_{tkl} 
+ 8 \lambda D_i K_{sj} D_l K^s_k \bigg)  \,,
\end{eqnarray}
 with the notation
\begin{equation}
X^{(i}\,_{j)} := \frac{1}{2} (X^i\,_j + X_j\,^i) \,, \qquad X^{[i}\,_{j]} := \frac{1}{2} (X^i\,_j - X_j\,^i) \,. \label{eq:defXij}
\end{equation}
We take Eq.~(\ref{eq:defXij}) as a definition, and it should be applied also when $X$ includes derivatives on~$r$, for instance $X^{(i} \dot{K}_{lj)} = \frac{1}{2}(X^i \dot{K}_{lj} + X_j \dot{K}^i_l)$.  

%% file: app_renorma.tex
\section{Technical details on Holographic Renormalization}
\label{sec:app_holo_renorm}

The renormalization procedure follows from an expansion of the four dimensional quantities in eigenfunctions of the dilatation operator

\begin{equation}
\delta_D = 2 \int d^4x \gamma_{ij} \frac{\delta}{\delta \gamma_{ij}} \,. 
\end{equation}
This expansion reads
\begin{eqnarray}
K^i_j &=& K_{(0)}\,^i_j + K_{(2)}\,^i_j + K_{(4)}\,^i_j + \tilde{K}_{(4)}\,^i_j \log e^{-2r} + \cdots \,, \label{eq:Kijexp} \\
A_i &=& A_{(0)}\,_i + A_{(2)}\,_i + \tilde{A}_{(2)}\,_i \log e^{-2r} + \cdots \,, \label{eq:Aiexp}
\end{eqnarray}
where
\begin{eqnarray}
&&\delta_D K_{(0)}\,^i_j = 0 \,, \qquad \delta_D K_{(2)}\,^i_j = -2 K_{(2)}\,^i_j \,, \nonumber \\
&& \delta_D K_{(4)}\,^i_j = -4 K_{(4)}\,^i_j - 2 \tilde{K}_{(4)}\,^i_j \,, \qquad \delta_D \tilde{K}_{(4)}\,^i_j = -4 \tilde{K}_{(4)}\,^i_j \,, \nonumber \\
&& \delta_D A_{(0)}\,_i = 0 \,, \qquad  \delta_D A_{(2)}\,_i  = -2 A_{(2)}\,_i - 2 \tilde{A}_{(2)}\,_i \,, \nonumber \\
&& \delta_D \tilde{A}_{(2)}\,_i = -2 \tilde{A}_{(2)}\,_i \,.
 \end{eqnarray}
Given the above expansion of the fields one has to solve the equations of motion in its Codazzi form, order by order in a recursive way. To do so one needs to identify the leading order in dilatation eigenvalues at which each term contributes. One has
\begin{eqnarray}
\gamma_{ij} &\sim& {\cal O}(-2)\,, \qquad \gamma^{ij} \sim {\cal O}(2)\,, \qquad\quad\; E_i \sim {\cal O}(2)\,, \quad\;\;\; \hat{F}_{ij} \sim {\cal O}(0)\,, \nonumber \\
\sqrt{-\gamma} &\sim& {\cal O}(-4)\,, \qquad  K^i_j \sim {\cal O}(0)\,, \qquad \hat R^i\,_{jkl} \sim {\cal O}(0)\,,  \qquad \nabla_i \sim {\cal O}(0) \,.
\end{eqnarray}
Note that for convenience of notation we define ${\cal O}(n)$ if the leading eigenvalue of the dilatation operator is $-n$.  In practice, in the renormalization procedure one needs to use the equations of motion Eqs.~(\ref{eq:Gb3}) and (\ref{eq:Gb1}) up to ${\cal O}(2)$ and ${\cal O}(4) + {\cal O}(\tilde{4})$ respectively. Up to ${\cal O}(0)$ they write
\begin{eqnarray}
&&0=K_{(0)}^2 - K_{(0)}\,^i_j K_{(0)}\,^j_i - 2\Lambda   \,,  \label{eq:Gb1order0}\\
&&0=\dot{K}_{(0)}\,^i_j + K_{(0)} K_{(0)}\,^i_j - \frac{2\Lambda}{(d-1)}\delta^i_j \,. \label{eq:Gb3order0}
\end{eqnarray}
Order ${\cal O}(2)$ writes
\begin{eqnarray}
&&0=2K_{(0)}K_{(2)} - 2 K_{(0)}\,^i_j K_{(2)}\,^j_i - \hat{R} \,, \label{eq:Gb1order2}  \\
&&0=\dot{K}^i_j|_{(2)} + K_{(0)} K_{(2)}\,^i_j + K_{(2)} K_{(0)}\,^i_j -\hat{R}^i_j \,, \label{eq:Gb3order2}
\end{eqnarray}
and finally orders ${\cal O}(4)$ and ${\cal O}(\tilde{4})$ for Eq.~(\ref{eq:Gb1}) write respectively
\begin{eqnarray}
&&0=2K_{(0)}K_{(4)} + K^2_{(2)} - 2 K_{(0)} \,^i_j K_{(4)}\,^j_i - K_{(2)}\,^i_j K_{(2)}\,^j_i + \frac{1}{4} \hat{F}_{(0)}\,_{ij} \hat{F}_{(0)}\,^{ij}  \,, \label{eq:Gb1order4}\\
&&0=2 \left( K_{(0)} \tilde{K}_{(4)} - K_{(0)}\,^i_j \tilde{K}_{(4)}\,^j_i \right)\log e^{-2r}  \,. \label{eq:Gb1order4tilde}
\end{eqnarray}
The derivative on $r$ can be computed by using 
\begin{equation}
\frac{d}{dr} = \int d^4x \dot{\gamma}_{km} \frac{\delta}{\delta \gamma_{km}} = 2 \int d^4x K^l_m \gamma_{lk} \frac{\delta}{\delta \gamma_{km}}  \,. \label{eq:dr}
\end{equation}
By inserting in this equation the expansion of $K^i_j$ given by Eq.~(\ref{eq:Kijexp}), one gets $d/dr \simeq \delta_D$ at the lowest order. Taking into account this, the computation of $K_{(0)}\,^i_j$ is trivial if one considers the definition of $K_{ij}$, i.e.
\begin{equation}
K_{(0)} \,_{ij} = \frac{1}{2} \dot\gamma_{ij} \big|_{(0)} = \frac{1}{2}\delta_D \gamma_{ij} = \gamma_{ij} \,. \label{eq:K0comp}
\end{equation}
Then the result up to ${\cal O}(0)$ is
\begin{eqnarray}
K_{(0)}\,^i_j = \delta^i_j \,, \qquad K_{(0)} = d \,. \label{eq:K0app}
\end{eqnarray}
Inserting this result into Eq.~(\ref{eq:Gb1order0}) or (\ref{eq:Gb3order0}) one arrives at the well known cosmological constant
\begin{equation}
\Lambda = \frac{d(d-1)}{2} \,.
\end{equation}
We have used in Eq.~(\ref{eq:Gb3order0}) that $\dot{K}_{(0)}\,^i_j = \delta_D K_{(0)}\,^i_j = 0$. The result for $K_{(2)}$ follows immediately from Eqs.~(\ref{eq:Gb1order2}) and (\ref{eq:K0app}),
\begin{equation}
K_{(2)} := P = \frac{\hat{R}}{2(d-1)} \,. \label{eq:P}
\end{equation}
In order to proceed with the computation of $K_{(2)}\,^i_j$ from Eq.~(\ref{eq:Gb3order2}), we should evaluate first $\dot{K}^i_j|_{(2)}$. Using the definition of $d/dr$ given by Eq.~(\ref{eq:dr}), it writes
\begin{eqnarray}
\dot{K}^i_j|_{(2)} &=& 2 \int d^4x K_{(0)}\,^l_m \gamma_{lk} \frac{\delta}{\delta\gamma_{km}} K_{(2)}\,^i_j + 2 \int d^4x K_{(2)}\,^l_m \gamma_{lk} \frac{\delta}{\delta\gamma_{km}} K_{(0)}\,^i_j \nonumber \\
&=& 2 \int d^4x \gamma_{km} \frac{\delta}{\delta\gamma_{km}} K_{(2)}\,^i_j = \delta_D K_{(2)}\,^i_j = -2 K_{(2)}\,^i_j \,. \label{eq:K2pijcomp}
\end{eqnarray}
Because $K_{(0)}\,^i_j$ is the Kronecker's delta, the second term after the first equality is zero, while the first one becomes the dilatation operator acting over $K_{(2)}\,^i_j$. Then one gets from Eq.~(\ref{eq:Gb3order2}) the result
\begin{equation}
K_{(2)}\,^i_j := P^i_j = \frac{1}{(d-2)} \left[ \hat{R}^i_j - P \delta^i_j \right] \,. \label{eq:Pij} 
\end{equation}
Note that the trace of $K_{(2)}\,^i_j$ agrees with Eq.~(\ref{eq:P}). Using all the results above it is straightforward to solve for orders ${\cal O}(4)$ and ${\cal O}(\tilde{4})$. From Eqs.~(\ref{eq:Gb1order4}) and (\ref{eq:Gb1order4tilde}) one gets respectively
\begin{eqnarray}
K_{(4)} &=& \frac{1}{2(d-1)} \bigg[ P^i_j P^j_i - P^2 - \frac{1}{4} \hat{F}_{(0)}\,_{ij} \hat{F}_{(0)}\,^{ij}  \bigg] \,, \label{eq:K4}\\
\tilde{K}_{(4)} &=& 0 \,. \label{eq:K4tilde}
\end{eqnarray}
 
In order to compute the counterterm for the on-shell action, besides the equations of motion an additional equation is needed. Following Ref.~\cite{Papadimitriou:2004ap}, one can introduce a covariant variable $\theta$ and write the on-shell action as
\begin{equation}
S_{on-shell} = \frac{1}{8\pi G} \int_\partial d^4x \sqrt{-\gamma} (K - \theta) \,. \label{eq:Sonshelltheta}
\end{equation}
Then computing $\dot{S}_{on-shell}$ from Eq.~(\ref{eq:Sonshelltheta}), and comparing it with the result obtained by using Eqs.~(\ref{eq:Sb1})-(\ref{eq:Sb3}), one gets the following equation
\begin{eqnarray}
0 &=& \dot{\theta} + K\theta -\frac{1}{(d-1)}\left( 2\Lambda + \frac{1}{2}E_i E^i + \frac{1}{4}\hat{F}_{ij}\hat{F}^{ij} \right) 
-\frac{2}{3}\kappa \epsilon^{ijkl} A_i E_j \hat{F}_{kl} \nonumber \\
&&-\frac{12\lambda}{(d-1)}\epsilon^{ijkl}\bigg[ 
 A_i\hat{R}^n\,_{mkl}D_n K^m_j 
+E_i K_{jm} D_k K^m_l
+\frac{1}{2}\hat{F}_{ik} K_{jm} \dot{K}^m_l  
\bigg]  \,. \label{eq:theta}
\end{eqnarray}
The variable $\theta$ admits also an expansion in eigenfunctions of $\delta_D$ of the form
\begin{equation}
\theta = \theta_{(0)} + \theta_{(2)} + \theta_{(4)} + \tilde\theta_{(4)}\log e^{-2r} + \cdots \,, \label{eq:thetaexpansion}
\end{equation}
where
\begin{eqnarray}
&&\delta_D \theta_{(0)} = 0 \,, \qquad \delta_D \theta_{(2)} = -2 \theta_{(2)} \,, \nonumber \\
&& \delta_D \theta_{(4)} = -4 \theta_{(4)} - 2 \tilde\theta_{(4)} \,, 
\qquad \delta_D \tilde\theta_{(4)} = -4 \tilde\theta_{(4)} \,.
\end{eqnarray}
Inserting expansion~(\ref{eq:thetaexpansion}) into Eq.~(\ref{eq:theta}), one gets the following identities
\begin{eqnarray}
&&0 = \dot{\theta}_{(0)} + K_{(0)}\theta_{(0)} - \frac{2\Lambda}{(d-1)} \,, \label{eq:thetaorder0} \\
&&0=\dot{\theta}|_{(2)} + K_{(2)}\theta_{(0)} + K_{(0)} \theta_{(2)} \,,  \label{eq:thetaorder2}\\
&&0=\dot{\theta}|_{(4)} + K_{(4)}\theta_{(0)} + K_{(2)} \theta_{(2)} + K_{(0)}\theta_{(4)}  - \frac{1}{4(d-1)}  \hat{F}_{(0)}\,_{ij} \hat{F}_{(0)}\,^{ij} \,, \label{eq:thetaorder4}\\
&&0=\dot{\theta}|_{(\tilde{4})}  + \left( \theta_{(0)} \tilde{K}_{(4)} + K_{(0)} \tilde{\theta}_{(4)} \right) \log e^{-2r} \,, \label{eq:thetaorder4tilde}
\end{eqnarray}
corresponding to orders ${\cal O}(0)$, ${\cal O}(2)$, ${\cal O}(4)$ and ${\cal O}(\tilde{4})$ respectively. Following the same procedure as shown in Eqs.~(\ref{eq:K0comp}) and (\ref{eq:K2pijcomp}), one gets
\begin{equation}
\dot{\theta}_{(0)} = 0 \,, \qquad \dot{\theta}_{(2)} = \delta_D \theta_{(2)} = -2\theta_{(2)} \,.
\end{equation}
At this point one can solve Eqs.~(\ref{eq:thetaorder0}) and (\ref{eq:thetaorder2}) to get
\begin{eqnarray}
\theta_{(0)} = 1 \,, \qquad \theta_{(2)} = \frac{P}{(2-d)} \,.  \label{eq:theta02}
\end{eqnarray}
Higher orders are a little bit more involved. Using the definition of
$d/dr$, then $\dot{\theta}|_{(4)}$ writes
\begin{eqnarray}
\dot{\theta}|_{(4)} &=& 2 \int d^4x K_{(0)}\,^l_m \gamma_{lk} \frac{\delta}{\delta\gamma_{km}} \theta_{(4)} 
+ 2 \int d^4x K_{(4)}\,^l_m \gamma_{lk} \frac{\delta}{\delta\gamma_{km}} \theta_{(0)} 
+2 \int d^4x K_{(2)}\,^l_m \gamma_{lk} \frac{\delta}{\delta\gamma_{km}} \theta_{(2)} \nonumber \\
&=& \delta_D \theta_{(4)} 
+ \frac{2}{(2-d)} \int d^4x P_{km} \frac{\delta}{\delta\gamma_{km}} P   \,. \label{eq:theta4pcomp}
\end{eqnarray}
Note that the second term after the first equality vanishes, while the first one writes in terms of $\delta_D$. To evaluate the last term at the r.h.s. of eq.~(\ref{eq:theta4pcomp}) we use
\begin{equation}
\delta \hat{R} = -\hat{R}^{km}\delta \gamma_{km} + D^k D^m \delta \gamma_{km} - \gamma^{km} D_l D^l \delta \gamma_{km} \,.
\end{equation}
After a straightforward computation, one gets
\begin{equation}
\dot{\theta}|_{(4)} = -4 \theta_{(4)} - 2\tilde\theta_{(4)} + \frac{1}{(d-1)(d-2)} \bigg[(d-2)P^i_j P^j_i + P^2 + D_i( D^i P  - D^j P^i_j) \bigg] \,. \label{eq:theta4dot}
\end{equation}
Inserting Eq.~(\ref{eq:theta4dot}) into Eq.~(\ref{eq:thetaorder4}) one can solve the latter, and the result is~\footnote{This result for $\tilde\theta_{(4)}$ includes a total derivative term which has not been computed in Ref.~\cite{Papadimitriou:2004ap}. To compute $\tilde\theta_{(4)}$, in this reference the authors derive the elegant relation $\tilde\theta_{(4)} = \frac{(d-1)}{2}K_{(4)} + \tilde{K}_{(4)}$. This identity is however valid modulo total derivative terms.}
\begin{equation}
\tilde\theta_{(4)} = \frac{1}{4} \bigg[ P^i_j P^j_i - P^2 - \frac{1}{4}  \hat{F}_{(0)}\,_{ij} \hat{F}_{(0)}\,^{ij} + \frac{1}{3} D_i\left( D^i P - D^j P^i_j \right)  \bigg] \,. \label{eq:theta4tilde}
\end{equation}
The computation of $\dot{\theta}|_{(\tilde{4})}$ follows in a similar way, and one gets $\dot{\theta}|_{(\tilde{4})} =  -4 \tilde\theta_{(4)} \log e^{-2r}$. By inserting it into Eq.~(\ref{eq:thetaorder4tilde}), this equation is trivially fulfilled.

The counterterm of the action can be read out from Eq.~(\ref{eq:Sonshelltheta}) by using $K$ and~$\theta$ computed up to order ${\cal O}(\tilde{4})$, i.e.
\begin{equation}
S_{ct} = -S_{on-shell} = -\frac{1}{8\pi G} \int_\partial d^4x \sqrt{-\gamma} \bigg[ (K_{(0)} - \theta_{(0)}) +  (K_{(2)} - \theta_{(2)})  + (\tilde{K}_{(4)} - \tilde{\theta}_{(4)}) \log e^{-2r} \bigg]  \,.
\end{equation}
From this equation and Eqs.~(\ref{eq:K0app}), (\ref{eq:P}), (\ref{eq:K4tilde}), (\ref{eq:theta02}) and (\ref{eq:theta4tilde}), one finally gets
\begin{eqnarray}
S_{ct} &=& - \frac{(d-1)}{8\pi G} \int_\partial d^4x \sqrt{-\gamma} \bigg[
1 + \frac{1}{(d-2)}P \nonumber \\
&&\qquad\qquad- \frac{1}{4(d-1)} \left( P^i_j P^j_i - P^2 -  \frac{1}{4} \hat{F}_{(0)}\,_{ij} \hat{F}_{(0)}\,^{ij} \right)\log e^{-2r} \bigg] \,.
\end{eqnarray}
The last term in Eq.~(\ref{eq:theta4tilde}) is a total derivative, and so it doesn't contribute to the action. As a remarkable fact we find that there is no contribution in the counterterm coming from the gauge-gravitational Chern-Simons term. This is because this term only contributes at higher orders. Indeed as explained above, in the renormalization procedure we use  Eqs.~(\ref{eq:Gb1}) and (\ref{eq:theta}) up to orders ${\cal O}(0)$, ${\cal O}(2)$, ${\cal O}(4)$ and ${\cal O}(\tilde{4})$, and Eq.~(\ref{eq:Gb3}) up to orders ${\cal O}(0)$ and ${\cal O}(2)$. We have explicitly checked that the $\lambda$ dependence starts contributing at ${\cal O}(6)$ in all these three equations.~\footnote{Note that $\dot{K}^i_j$ and $\dot{\theta}$ induce terms proportional to $\lambda$. Up to order ${\cal O}(4) + {\cal O}(\tilde{4})$ these operators write $\dot{K}^i_j|_{(4)+(\tilde{4})} = -4 K_{(4)}\,^i_j + \dots$, and $\dot{\theta}|_{(4)+(\tilde{4})} = -4 \theta_{(4)} + \dots$, where the dots indicate extra terms which are $\lambda$-independent. The only $\lambda$-dependence could appear in $K_{(4)}\,^i_j$ and $\theta_{(4)}$, but these contributions are precisely cancelled by other terms in Eqs.~(\ref{eq:Gb3}) and (\ref{eq:theta}) respectively, so that these equations become $\lambda$-dependent only at ${\cal O}(6)$ and higher.} This means that the gauge-gravitational Chern-Simons term does not induce new divergences, and so the renormalization is not modified by it.

%% file: equations_shear.tex
\section{Equations of motion for the shear sector}
\label{eq_shear}
These are the complete linearized set of six dynamical equations of motion
\begin{eqnarray}
\label{eq_As}\nonumber 0&=& B_\alpha'' ( u )+\frac { f' ( u ) }{f(u)}B_\alpha'( u )+ \frac {b^2  }{uf( u )^2 }\left(w^2-  f( u) 
{k}^{2} \right)B_\alpha( u) -  \frac{ h^{\alpha'}_t ( u
 )}{f(u)} \\
 &&+ik\epsilon_{\alpha\beta}\left(\ \frac{3}{u f(u)} \bar\lambda \left( \frac{2}{3a} (f(u)-1)+ u^3 \right) h_t^{\beta'}(u)+\bar \kappa \frac{B_\beta(u)}{f(u)} \right) \,,\\
\label{eq_Hts}\nonumber 0&=&  h^{\alpha''}_t(u) - \frac{h^{\alpha'}_t(u)}{u} -\frac {b^2}{
uf( u )}\left(k^2 h^\alpha_t(u)+ h^\alpha_y \left( u \right) wk \right)
- 3a u B'_\alpha(u) \\
\nonumber &&i\bar\lambda k \epsilon_{\alpha\beta}\left[\left(24a u^3-6(1-f(u))\right)\frac {B_\beta(u)}{u}+(9a u^3-6(1-f(u)))B'_\beta(u)\right.\\
 &&\left.+2 u (uh^{\beta'}_t ( u ) )'  -\frac {  2u  b^2}{  f ( u) }\left( h_y^\beta( u ) w k+h_t^\beta( u ) k^2\right)\right] \,,\\
\label{eq_Hys}\nonumber 0&=& h^{\alpha''}_y(u)+ \frac{\left(f/u\right)'}{f/u}h^{\alpha'}_y(u)+\frac{b^2}{u f(u)^2}\left(w^2h_y^\alpha(u)+wk h_t^\alpha(u)\right)+2uik\bar{\lambda}\epsilon_{\alpha\beta}\left[u  h^{\beta''}_y(u)\right.\\
 && \left.+\left(9f(u)-6+3a u^3\right)\frac{h^{\beta'}_y(u)  }{f(u)}+\frac{ b^2 }{f(u)^2}\left(wk h_t^\beta(u)+w^2 h_y^\beta(u)\right)\right] \,,
\end{eqnarray}

and two constraints for the fluctuations at $w ,k\neq 0$ 
\begin{eqnarray}
\label{constraints}\nonumber 0&=&  w \left(h^{\alpha'}_t(u)-3a u B_\alpha(u)\right)+f(u)kh^{\alpha'}_y(u)+i k\bar\lambda\epsilon_{\alpha\beta}\left[2u^2 \left(w h^{\beta'}_t+ f( u )  k h^{\beta'}_y (u)\right)\right.\\
 && \left. +\left(9a u^3-6(1-f(u))\right)B_\beta(u)\right] \,.
\end{eqnarray}

%% file: sol_w_0.tex
\section{Solutions at zero frequency}
\label{sol_w_0}
We write in this appendix the solutions for the system
(\ref{htx1})-(\ref{bx1}). These functions depend explicitly on the
boundary sources $\tilde H^\alpha$ and $\tilde B_\alpha$, and the
anomalous parameters $\bar \kappa,\bar\lambda$. Switching off $\bar
\lambda$ we get the same system obtained in \cite{Amado:2011zx}
{\small
\begin{eqnarray}
\nonumber h^\alpha_t(u) &=& \tilde H^\alpha f(u) - \frac{i k \bar\kappa\epsilon_{\alpha\beta} (u-1)a }{2 (1+4 a)^{3/2}}\left[
\left. (1+4 a)^{3/2} u^2 \tilde H^\beta\right.\right.\\
\nonumber&&\left. 3 \left(\sqrt{1+4 a} u (2 a u-1)+2 \left(1+u-a u^2\right) \text{ArcCoth}\left[\frac{2+u}{\sqrt{1+4 a} u}\right]\right)\tilde B_\beta\right]
\\
\nonumber&&k i\bar\lambda \epsilon_{\alpha\beta}(u-1) \left[\tilde B_\beta  \left(-\frac{3 i (u+1)(1+a) \pi }{2  a }+\frac{3   (1+a (5+a)) u}{ (1+4 a)}+\frac{  \left(5+21 a+2 a^3\right) u^2}{ (1+4 a)}\right.\right.\\
\nonumber&&\left.\left.+\frac{3}{2}i  (1+a) \pi  u^2-6   a u^3
-\frac{3i f(u) (1+a (7+2 a (7+a))) \text{ArcCoth}\left[\frac{2+u}{\sqrt{1+4 a} u}\right]}{(u-1)(-1-4 a)^{3/2}  a}+\right.\right.\\
 &&\left.\left.-\frac{3 f(u) (1+a) }{2a(u-1)} \text{Log}\left[-1-u+a u^2\right]\right)+\right.\\
\nonumber&&\left.\tilde H^\beta  \left(-\frac{2i (u+1) (1+a)^2 \pi }{ a^2 }+\frac{2   (1+a) (2+a (7+2 a)) u}{ a (1+4 a) }+\right.\right.\\
\nonumber&&\left.\left. +\frac{  (4+a (25+a (39+a (-5+4 a)))) u^2}{ a (1+4 a) }+\frac{2i  (1+a)^2 \pi  u^2}{ a }+\right.\right.\\
\nonumber&&\left.\left.   u^3(1-5 a -6au)+\right.\right.\\
\nonumber&&\left.\left. -\frac{4 i f(u) (1+a) (1+2 a) (1+a (5+a)) \text{ArcCoth}\left[\frac{2+u}{\sqrt{1+4 a} u}\right]}{(u-1)(-1-4 a)^{3/2}  a^2}+\right.\right.\\
\nonumber&&\left.\left.-\frac{2  f(u) (1+a)^2 \text{Log}\left[-1-u+a u^2\right]}{ (u-1)a^2}\right)\right]\,,
\end{eqnarray}
\begin{eqnarray}
\nonumber B_\alpha(u)&=& \tilde B_\alpha+\tilde H^\alpha u -i\frac{k\bar\kappa\epsilon_{\alpha\beta}}{2 (1+4 a)^{3/2}} \left(\tilde H^\beta u\left(  1+4 a   \right)^{3/2}+\right.\\
\nonumber&&\left.\tilde B_\beta \left(6 a \sqrt{1+4 a}  u-2 (-2+a (-2+3 u)) \text{ArcCoth}\left[\frac{2+u}{\sqrt{1+4 a} u}\right]\right)\right)\\
\nonumber&&+i k \bar\lambda \epsilon_{\alpha\beta}\left[\tilde B_\beta \left(-\frac{i (1+a)^2  \pi }{ a^2 }+\frac{2   (1+a) (1+a (5+a))  u}{ a (1+4 a )}+\frac{3 i (1+a)  \pi  u}{2  a }\right.\right.\\
\nonumber&&\left.\left. -3 u^2-\frac{i (1+a (7+2 a (7+a))) (-2+a (-2+3 u)) \text{ArcCoth}\left[\frac{2+u}{\sqrt{1+4 a} u}\right]}{(-1-4 a)^{3/2}  a^2}\right.\right.\\
&&\left.\left.-\frac{ (1+a) (-2+a (-2+3 u)) \text{Log}\left[-1-u+a u^2\right]}{2 a^2}\right)\right.\\
\nonumber && \left.+\tilde H^\beta \left(-\frac{4 i (1+a)^3  \pi }{3  a^3 }+\frac{  (8+a (48+a (84+a (29+12 a))))  u}{3  a^2 (1+4 a )}+\right.\right.\\
\nonumber&&\left.\left.+\frac{2 i (1+a)^2  \pi  u}{ a^2 }-\frac{2   (1+a)  u^2}{ a }-3 u^3\right.\right.\\
\nonumber&&\left.\left.-\frac{4 i (1+a) (1+2 a) (1+a (5+a)) (-2+a (-2+3 u)) \text{ArcCoth}\left[\frac{2+u}{\sqrt{1+4 a} u}\right]}{3 (-1-4 a)^{3/2}  a^3}+\right.\right.\\
\nonumber&&\left.\left.-\frac{2 (1+a)^2 (-2+a (-2+3 u)) \text{Log}\left[-1-u+a u^2\right]}{3  a^3}\right)\right]\,.
\end{eqnarray}
}

%% file: HoloGravAnomal.bbl
\providecommand{\href}[2]{#2}\begingroup\raggedright\begin{thebibliography}{10}

\bibitem{Adler:1969gk}
S.~L. Adler, {\it {Axial vector vertex in spinor electrodynamics}},  {\em
  Phys.Rev.} {\bf 177} (1969) 2426--2438.

\bibitem{Bell:1969ts}
J.~S. Bell and R.~Jackiw, {\it {A PCAC puzzle: $\pi\_0 \to \gamma \gamma$ in
  the sigma model}},  {\em Nuovo Cim.} {\bf A60} (1969) 47--61.

\bibitem{Delbourgo:1972xb}
R.~Delbourgo and A.~Salam, {\it {The gravitational correction to pcac}},  {\em
  Phys.Lett.} {\bf B40} (1972) 381--382.

\bibitem{Eguchi:1976db}
T.~Eguchi and P.~G. Freund, {\it {Quantum Gravity and World Topology}},  {\em
  Phys.Rev.Lett.} {\bf 37} (1976) 1251.

\bibitem{AlvarezGaume:1983ig}
L.~Alvarez-Gaume and E.~Witten, {\it {Gravitational Anomalies}},  {\em
  Nucl.Phys.} {\bf B234} (1984) 269.

\bibitem{Fukushima:2008xe}
K.~Fukushima, D.~E. Kharzeev, and H.~J. Warringa, {\it {The Chiral Magnetic
  Effect}},  {\em Phys. Rev.} {\bf D78} (2008) 074033,
  [\href{http://xxx.lanl.gov/abs/0808.3382}{{\tt arXiv:0808.3382}}].

\bibitem{Erdmenger:2008rm}
J.~Erdmenger, M.~Haack, M.~Kaminski, and A.~Yarom, {\it {Fluid dynamics of
  R-charged black holes}},  {\em JHEP} {\bf 01} (2009) 055,
  [\href{http://xxx.lanl.gov/abs/0809.2488}{{\tt arXiv:0809.2488}}].

\bibitem{Banerjee:2008th}
N.~Banerjee {\em et.~al.}, {\it {Hydrodynamics from charged black branes}},
  {\em JHEP} {\bf 01} (2011) 094,
  [\href{http://xxx.lanl.gov/abs/0809.2596}{{\tt arXiv:0809.2596}}].

\bibitem{Son:2009tf}
D.~T. Son and P.~Surowka, {\it {Hydrodynamics with Triangle Anomalies}},  {\em
  Phys. Rev. Lett.} {\bf 103} (2009) 191601,
  [\href{http://xxx.lanl.gov/abs/0906.5044}{{\tt arXiv:0906.5044}}].

\bibitem{Vilenkin:1995um}
A.~Vilenkin, {\it {Parity nonconservation and neutrino transport in magnetic
  fields}},  {\em Astrophys.J.} {\bf 451} (1995) 700--702.

\bibitem{Vilenkin:1980ft}
A.~Vilenkin, {\it {Cancellation of equilibrium parity violating currents}},
  {\em Phys.Rev.} {\bf D22} (1980) 3067--3079.

\bibitem{Vilenkin:1980zv}
A.~Vilenkin, {\it {Quntum field theory at finite temperature in a rotating
  system}},  {\em Phys.Rev.} {\bf D21} (1980) 2260--2269.

\bibitem{Vilenkin:1978hb}
A.~Vilenkin, {\it {Parity Violating Currents in Thermal Radiation}},  {\em
  Phys.Lett.} {\bf B80} (1978) 150--152.

\bibitem{Giovannini:1997eg}
M.~Giovannini and M.~Shaposhnikov, {\it {Primordial hypermagnetic fields and
  triangle anomaly}},  {\em Phys.Rev.} {\bf D57} (1998) 2186--2206,
  [\href{http://xxx.lanl.gov/abs/hep-ph/9710234}{{\tt hep-ph/9710234}}].

\bibitem{Alekseev:1998ds}
A.~Y. Alekseev, V.~V. Cheianov, and J.~Fr{\"o}hlich, {\it {Universality of
  transport properties in equilibrium, Goldstone theorem and chiral anomaly}},
  {\em Phys. Rev. Lett.} {\bf 81} (1998) 3503--3506,
  [\href{http://xxx.lanl.gov/abs/cond-mat/9803346}{{\tt cond-mat/9803346}}].

\bibitem{Kharzeev:2007jp}
D.~E. Kharzeev, L.~D. McLerran, and H.~J. Warringa, {\it {The effects of
  topological charge change in heavy ion collisions: 'Event by event P and CP
  violation'}},  {\em Nucl. Phys.} {\bf A803} (2008) 227--253,
  [\href{http://xxx.lanl.gov/abs/0711.0950}{{\tt arXiv:0711.0950}}].

\bibitem{Newman:2005hd}
G.~M. Newman, {\it {Anomalous hydrodynamics}},  {\em JHEP} {\bf 01} (2006) 158,
  [\href{http://xxx.lanl.gov/abs/hep-ph/0511236}{{\tt hep-ph/0511236}}].

\bibitem{Kharzeev:2007tn}
D.~Kharzeev and A.~Zhitnitsky, {\it {Charge separation induced by P-odd bubbles
  in QCD matter}},  {\em Nucl. Phys.} {\bf A797} (2007) 67--79,
  [\href{http://xxx.lanl.gov/abs/0706.1026}{{\tt arXiv:0706.1026}}].

\bibitem{Yee:2009vw}
H.-U. Yee, {\it {Holographic Chiral Magnetic Conductivity}},  {\em JHEP} {\bf
  11} (2009) 085, [\href{http://xxx.lanl.gov/abs/0908.4189}{{\tt
  arXiv:0908.4189}}].

\bibitem{Torabian:2009qk}
M.~Torabian and H.-U. Yee, {\it {Holographic nonlinear hydrodynamics from
  AdS/CFT with multiple/non-Abelian symmetries}},  {\em JHEP} {\bf 08} (2009)
  020, [\href{http://xxx.lanl.gov/abs/0903.4894}{{\tt arXiv:0903.4894}}].

\bibitem{Rebhan:2009vc}
A.~Rebhan, A.~Schmitt, and S.~A. Stricker, {\it {Anomalies and the chiral
  magnetic effect in the Sakai- Sugimoto model}},  {\em JHEP} {\bf 01} (2010)
  026, [\href{http://xxx.lanl.gov/abs/0909.4782}{{\tt arXiv:0909.4782}}].

\bibitem{Brits:2010pw}
L.~Brits and J.~Charbonneau, {\it {A Constraint-Based Approach to the Chiral
  Magnetic Effect}},  \href{http://xxx.lanl.gov/abs/1009.4230}{{\tt
  arXiv:1009.4230}}.

\bibitem{Gynther:2010ed}
A.~Gynther, K.~Landsteiner, F.~Pena-Benitez, and A.~Rebhan, {\it {Holographic
  Anomalous Conductivities and the Chiral Magnetic Effect}},  {\em JHEP} {\bf
  02} (2011) 110, [\href{http://xxx.lanl.gov/abs/1005.2587}{{\tt
  arXiv:1005.2587}}].

\bibitem{Gorsky:2010xu}
A.~Gorsky, P.~N. Kopnin, and A.~V. Zayakin, {\it {On the Chiral Magnetic Effect
  in Soft-Wall AdS/QCD}},  {\em Phys. Rev.} {\bf D83} (2011) 014023,
  [\href{http://xxx.lanl.gov/abs/1003.2293}{{\tt arXiv:1003.2293}}].

\bibitem{Kalaydzhyan:2011vx}
T.~Kalaydzhyan and I.~Kirsch, {\it {Fluid/gravity model for the chiral magnetic
  effect}},  {\em Phys. Rev. Lett.} {\bf 106} (2011) 211601,
  [\href{http://xxx.lanl.gov/abs/1102.4334}{{\tt arXiv:1102.4334}}].

\bibitem{Hoyos:2011us}
C.~Hoyos, T.~Nishioka, and A.~O'Bannon, {\it {A Chiral Magnetic Effect from
  AdS/CFT with Flavor}},  \href{http://xxx.lanl.gov/abs/1106.4030}{{\tt
  arXiv:1106.4030}}.

\bibitem{Eling:2010hu}
C.~Eling, Y.~Neiman, and Y.~Oz, {\it {Holographic Non-Abelian Charged
  Hydrodynamics from the Dynamics of Null Horizons}},  {\em JHEP} {\bf 12}
  (2010) 086, [\href{http://xxx.lanl.gov/abs/1010.1290}{{\tt
  arXiv:1010.1290}}].

\bibitem{Buividovich:2009wi}
P.~V. Buividovich, M.~N. Chernodub, E.~V. Luschevskaya, and M.~I. Polikarpov,
  {\it {Numerical evidence of chiral magnetic effect in lattice gauge theory}},
   {\em Phys. Rev.} {\bf D80} (2009) 054503,
  [\href{http://xxx.lanl.gov/abs/0907.0494}{{\tt arXiv:0907.0494}}].

\bibitem{Abramczyk:2009gb}
M.~Abramczyk, T.~Blum, G.~Petropoulos, and R.~Zhou, {\it {Chiral magnetic
  effect in 2+1 flavor QCD+QED}},  {\em PoS} {\bf LAT2009} (2009) 181,
  [\href{http://xxx.lanl.gov/abs/0911.1348}{{\tt arXiv:0911.1348}}].

\bibitem{Yamamoto:2011gk}
A.~Yamamoto, {\it {Chiral magnetic effect in lattice QCD with chiral chemical
  potential}},  \href{http://xxx.lanl.gov/abs/1105.0385}{{\tt
  arXiv:1105.0385}}.

\bibitem{Newman:2005as}
G.~M. Newman and D.~T. Son, {\it {Response of strongly-interacting matter to
  magnetic field: Some exact results}},  {\em Phys. Rev.} {\bf D73} (2006)
  045006, [\href{http://xxx.lanl.gov/abs/hep-ph/0510049}{{\tt
  hep-ph/0510049}}].

\bibitem{Kharzeev:2010gd}
D.~E. Kharzeev and H.-U. Yee, {\it {Chiral Magnetic Wave}},  {\em Phys. Rev.}
  {\bf D83} (2011) 085007, [\href{http://xxx.lanl.gov/abs/1012.6026}{{\tt
  arXiv:1012.6026}}].

\bibitem{Kharzeev:2010gr}
D.~E. Kharzeev and D.~T. Son, {\it {Testing the chiral magnetic and chiral
  vortical effects in heavy ion collisions}},  {\em Phys. Rev. Lett.} {\bf 106}
  (2011) 062301, [\href{http://xxx.lanl.gov/abs/1010.0038}{{\tt
  arXiv:1010.0038}}].

\bibitem{KerenZur:2010zw}
B.~Keren-Zur and Y.~Oz, {\it {Hydrodynamics and the Detection of the QCD Axial
  Anomaly in Heavy Ion Collisions}},  {\em JHEP} {\bf 06} (2010) 006,
  [\href{http://xxx.lanl.gov/abs/1002.0804}{{\tt arXiv:1002.0804}}].

\bibitem{Asakawa:2010bu}
M.~Asakawa, A.~Majumder, and B.~Muller, {\it {Electric Charge Separation in
  Strong Transient Magnetic Fields}},  {\em Phys. Rev.} {\bf C81} (2010)
  064912, [\href{http://xxx.lanl.gov/abs/1003.2436}{{\tt arXiv:1003.2436}}].

\bibitem{Ajitanand:2010rc}
N.~N. Ajitanand, R.~A. Lacey, A.~Taranenko, and J.~M. Alexander, {\it {A new
  method for the experimental study of topological effects in the quark-gluon
  plasma}},  {\em Phys. Rev.} {\bf C83} (2011) 011901,
  [\href{http://xxx.lanl.gov/abs/1009.5624}{{\tt arXiv:1009.5624}}].

\bibitem{abelev:2009uh}
{\bf STAR} Collaboration, B.~I. Abelev {\em et.~al.}, {\it {Azimuthal
  Charged-Particle Correlations and Possible Local Strong Parity Violation}},
  {\em Phys. Rev. Lett.} {\bf 103} (2009) 251601,
  [\href{http://xxx.lanl.gov/abs/0909.1739}{{\tt arXiv:0909.1739}}].

\bibitem{abelev:2009txa}
{\bf STAR} Collaboration, B.~I. Abelev {\em et.~al.}, {\it {Observation of
  charge-dependent azimuthal correlations and possible local strong parity
  violation in heavy ion collisions}},  {\em Phys. Rev.} {\bf C81} (2010)
  054908, [\href{http://xxx.lanl.gov/abs/0909.1717}{{\tt arXiv:0909.1717}}].

\bibitem{Kharzeev:2009pj}
D.~E. Kharzeev and H.~J. Warringa, {\it {Chiral Magnetic conductivity}},  {\em
  Phys. Rev.} {\bf D80} (2009) 034028,
  [\href{http://xxx.lanl.gov/abs/0907.5007}{{\tt arXiv:0907.5007}}].

\bibitem{Hou:2011ze}
D.~Hou, H.~Liu, and H.-c. Ren, {\it {Some Field Theoretic Issues Regarding the
  Chiral Magnetic Effect}},  {\em JHEP} {\bf 05} (2011) 046,
  [\href{http://xxx.lanl.gov/abs/1103.2035}{{\tt arXiv:1103.2035}}].

\bibitem{Amado:2011zx}
I.~Amado, K.~Landsteiner, and F.~Pena-Benitez, {\it {Anomalous transport
  coefficients from Kubo formulas in Holography}},  {\em JHEP} {\bf 1105}
  (2011) 081, [\href{http://xxx.lanl.gov/abs/1102.4577}{{\tt
  arXiv:1102.4577}}].

\bibitem{LeBellac:1991cq}
M.~Le~Bellac, {\it {Equilibrium and Non-Equilibrium Statistical
  Thermodynamics}}, . Cambridge University Press (2006).

\bibitem{Kharzeev:2011ds}
D.~E. Kharzeev and H.-U. Yee, {\it {Anomalies and time reversal invariance in
  relativistic hydrodynamics: the second order and higher dimensional
  formulations}},  \href{http://xxx.lanl.gov/abs/1105.6360}{{\tt
  arXiv:1105.6360}}.

\bibitem{Landsteiner:2011cp}
K.~Landsteiner, E.~Megias, and F.~Pena-Benitez, {\it {Gravitational Anomaly and
  Transport}},  \href{http://xxx.lanl.gov/abs/1103.5006}{{\tt
  arXiv:1103.5006}}. Physical Review Letters, in press.

\bibitem{Sadofyev:2010pr}
A.~V. Sadofyev and M.~V. Isachenkov, {\it {The chiral magnetic effect in
  hydrodynamical approach}},  {\em Phys. Lett.} {\bf B697} (2011) 404--406,
  [\href{http://xxx.lanl.gov/abs/1010.1550}{{\tt arXiv:1010.1550}}].

\bibitem{Neiman:2010zi}
Y.~Neiman and Y.~Oz, {\it {Relativistic Hydrodynamics with General Anomalous
  Charges}},  {\em JHEP} {\bf 03} (2011) 023,
  [\href{http://xxx.lanl.gov/abs/1011.5107}{{\tt arXiv:1011.5107}}].

\bibitem{Bhattacharya:2011tr}
J.~Bhattacharya, S.~Bhattacharyya, S.~Minwalla, and A.~Yarom, {\it {A Theory of
  first order dissipative superfluid dynamics}},
  \href{http://xxx.lanl.gov/abs/1105.3733}{{\tt arXiv:1105.3733}}.

\bibitem{Lin:2011mr}
S.~Lin, {\it {On the anomalous superfluid hydrodynamics}},
  \href{http://xxx.lanl.gov/abs/1104.5245}{{\tt arXiv:1104.5245}}.

\bibitem{Neiman:2011mj}
Y.~Neiman and Y.~Oz, {\it {Anomalies in Superfluids and a Chiral Electric
  Effect}},  \href{http://xxx.lanl.gov/abs/1106.3576}{{\tt arXiv:1106.3576}}.

\bibitem{Loganayagam:2011mu}
R.~Loganayagam, {\it {Anomaly Induced Transport in Arbitrary Dimensions}},
  \href{http://xxx.lanl.gov/abs/1106.0277}{{\tt arXiv:1106.0277}}.

\bibitem{Maldacena:1997re}
J.~M. Maldacena, {\it {The large N limit of superconformal field theories and
  supergravity}},  {\em Adv. Theor. Math. Phys.} {\bf 2} (1998) 231--252,
  [\href{http://xxx.lanl.gov/abs/hep-th/9711200}{{\tt hep-th/9711200}}].

\bibitem{Gubser:1998bc}
S.~S. Gubser, I.~R. Klebanov, and A.~M. Polyakov, {\it {Gauge theory
  correlators from non-critical string theory}},  {\em Phys. Lett.} {\bf B428}
  (1998) 105--114, [\href{http://xxx.lanl.gov/abs/hep-th/9802109}{{\tt
  hep-th/9802109}}].

\bibitem{Witten:1998qj}
E.~Witten, {\it {Anti-de Sitter space and holography}},  {\em Adv. Theor. Math.
  Phys.} {\bf 2} (1998) 253--291,
  [\href{http://xxx.lanl.gov/abs/hep-th/9802150}{{\tt hep-th/9802150}}].

\bibitem{Aharony:1999ti}
O.~Aharony, S.~S. Gubser, J.~M. Maldacena, H.~Ooguri, and Y.~Oz, {\it {Large N
  field theories, string theory and gravity}},  {\em Phys. Rept.} {\bf 323}
  (2000) 183--386, [\href{http://xxx.lanl.gov/abs/hep-th/9905111}{{\tt
  hep-th/9905111}}].

\bibitem{Andreev:2006ct}
O.~Andreev and V.~I. Zakharov, {\it {Heavy-quark potentials and AdS/QCD}},
  {\em Phys. Rev.} {\bf D74} (2006) 025023,
  [\href{http://xxx.lanl.gov/abs/hep-ph/0604204}{{\tt hep-ph/0604204}}].

\bibitem{Galow:2009kw}
B.~Galow, E.~Megias, J.~Nian, and H.~J. Pirner, {\it {Phenomenology of AdS/QCD
  and Its Gravity Dual}},  {\em Nucl. Phys.} {\bf B834} (2010) 330--362,
  [\href{http://xxx.lanl.gov/abs/0911.0627}{{\tt arXiv:0911.0627}}].

\bibitem{Gursoy:2008za}
U.~Gursoy, E.~Kiritsis, L.~Mazzanti, and F.~Nitti, {\it {Holography and
  Thermodynamics of 5D Dilaton-gravity}},  {\em JHEP} {\bf 05} (2009) 033,
  [\href{http://xxx.lanl.gov/abs/0812.0792}{{\tt arXiv:0812.0792}}].

\bibitem{Megias:2010ku}
E.~Megias, H.~J. Pirner, and K.~Veschgini, {\it {QCD-Thermodynamics using 5-dim
  Gravity}},  {\em Phys. Rev.} {\bf D83} (2011) 056003,
  [\href{http://xxx.lanl.gov/abs/1009.2953}{{\tt arXiv:1009.2953}}].

\bibitem{Veschgini:2010ws}
K.~Veschgini, E.~Megias, and H.~J. Pirner, {\it {Trouble Finding the Optimal
  AdS/QCD}},  {\em Phys. Lett.} {\bf B696} (2011) 495--498,
  [\href{http://xxx.lanl.gov/abs/1009.4639}{{\tt arXiv:1009.4639}}].

\bibitem{Jackiw:2003pm}
R.~Jackiw and S.~Pi, {\it {Chern-Simons modification of general relativity}},
  {\em Phys.Rev.} {\bf D68} (2003) 104012,
  [\href{http://xxx.lanl.gov/abs/gr-qc/0308071}{{\tt gr-qc/0308071}}].

\bibitem{Alexander:2009tp}
S.~Alexander and N.~Yunes, {\it {Chern-Simons Modified General Relativity}},
  {\em Phys.Rept.} {\bf 480} (2009) 1--55,
  [\href{http://xxx.lanl.gov/abs/0907.2562}{{\tt arXiv:0907.2562}}].

\bibitem{Saremi:2011ab}
O.~Saremi and D.~T. Son, {\it {Hall viscosity from gauge/gravity duality}},
  \href{http://xxx.lanl.gov/abs/1103.4851}{{\tt arXiv:1103.4851}}.

\bibitem{Delsate:2011qp}
T.~Delsate, V.~Cardoso, and P.~Pani, {\it {Anti de Sitter black holes and
  branes in dynamical Chern-Simons gravity: perturbations, stability and the
  hydrodynamic modes}},  {\em JHEPA,1106,055.2011} {\bf 1106} (2011) 055,
  [\href{http://xxx.lanl.gov/abs/1103.5756}{{\tt arXiv:1103.5756}}].

\bibitem{Bertlmann:1996xk}
R.~A. Bertlmann, {\it {Anomalies in quantum field theory}}, . Oxford, UK:
  Clarendon (1996) 566 p. (International series of monographs on physics: 91).

\bibitem{Martelli:2002sp}
D.~Martelli and W.~Mueck, {\it {Holographic renormalization and Ward identities
  with the Hamilton-Jacobi method}},  {\em Nucl.Phys.} {\bf B654} (2003)
  248--276, [\href{http://xxx.lanl.gov/abs/hep-th/0205061}{{\tt
  hep-th/0205061}}].

\bibitem{Papadimitriou:2004ap}
I.~Papadimitriou and K.~Skenderis, {\it {AdS / CFT correspondence and
  geometry}},  \href{http://xxx.lanl.gov/abs/hep-th/0404176}{{\tt
  hep-th/0404176}}.

\bibitem{Yee:2011yn}
H.-U. Yee and I.~Zahed, {\it {Holographic two dimensional QCD and Chern-Simons
  term}},  \href{http://xxx.lanl.gov/abs/1103.6286}{{\tt arXiv:1103.6286}}.

\bibitem{Kraus:2005zm}
P.~Kraus and F.~Larsen, {\it {Holographic gravitational anomalies}},  {\em
  JHEP} {\bf 0601} (2006) 022,
  [\href{http://xxx.lanl.gov/abs/hep-th/0508218}{{\tt hep-th/0508218}}].

\bibitem{Grumiller:2008ie}
D.~Grumiller, R.~B. Mann, and R.~McNees, {\it {Dirichlet boundary value problem
  for Chern-Simons modified gravity}},  {\em Phys.Rev.} {\bf D78} (2008)
  081502, [\href{http://xxx.lanl.gov/abs/0803.1485}{{\tt arXiv:0803.1485}}].

\bibitem{Clark:2010fs}
T.~Clark, S.~Love, and T.~ter Veldhuis, {\it {Holographic Currents and
  Chern-Simons Terms}},  {\em Phys.Rev.} {\bf D82} (2010) 106004,
  [\href{http://xxx.lanl.gov/abs/1006.2400}{{\tt arXiv:1006.2400}}].

\bibitem{Son:2002sd}
D.~T. Son and A.~O. Starinets, {\it {Minkowski-space correlators in AdS/CFT
  correspondence: Recipe and applications}},  {\em JHEP} {\bf 09} (2002) 042,
  [\href{http://xxx.lanl.gov/abs/hep-th/0205051}{{\tt hep-th/0205051}}].

\bibitem{Herzog:2002pc}
C.~P. Herzog and D.~T. Son, {\it {Schwinger-Keldysh propagators from AdS/CFT
  correspondence}},  {\em JHEP} {\bf 03} (2003) 046,
  [\href{http://xxx.lanl.gov/abs/hep-th/0212072}{{\tt hep-th/0212072}}].

\bibitem{Kaminski:2009dh}
M.~Kaminski, K.~Landsteiner, J.~Mas, J.~P. Shock, and J.~Tarrio, {\it
  {Holographic Operator Mixing and Quasinormal Modes on the Brane}},  {\em
  JHEP} {\bf 02} (2010) 021, [\href{http://xxx.lanl.gov/abs/0911.3610}{{\tt
  arXiv:0911.3610}}].

\bibitem{Amado:2009ts}
I.~Amado, M.~Kaminski, and K.~Landsteiner, {\it {Hydrodynamics of Holographic
  Superconductors}},  {\em JHEP} {\bf 0905} (2009) 021,
  [\href{http://xxx.lanl.gov/abs/0903.2209}{{\tt 0903.2209}}].

\bibitem{Bonora:2011mf}
L.~Bonora, M.~Cvitan, P.~D. Prester, S.~Pallua, and I.~Smolic, {\it
  {Gravitational Chern-Simons Lagrangian terms and spherically symmetric
  spacetimes}},  \href{http://xxx.lanl.gov/abs/1105.4792}{{\tt
  arXiv:1105.4792}}.

\bibitem{Bhattacharyya:2008jc}
S.~Bhattacharyya, V.~E. Hubeny, S.~Minwalla, and M.~Rangamani, {\it {Nonlinear
  Fluid Dynamics from Gravity}},  {\em JHEP} {\bf 0802} (2008) 045,
  [\href{http://xxx.lanl.gov/abs/0712.2456}{{\tt arXiv:0712.2456}}].

\bibitem{workinprogress}
K.~Landsteiner, E.~Megias, and F.~Pena-Benitez, {\it {Work in progress}}, .

\end{thebibliography}\endgroup
